\documentclass[final,3p,times,twocolumn]{elsarticle}

\usepackage{graphicx}
\usepackage{color}

\usepackage{amssymb}
\usepackage{amsthm}
\usepackage{amsmath}
\usepackage{subfigure}
\usepackage{tikz}

\usepackage{romannum} 

\usepackage{caption}    

\renewcommand{\vec}[1]{\mathbf{#1}}
\newcommand{\grvec}[1]{\boldsymbol{#1}}

\newcommand{\llrrvline}[1]{
	|\mkern-2mu|#1|\mkern-2mu|}

\DeclareMathOperator*{\argmin}{arg\,min} 
\DeclareMathOperator*{\argmax}{arg\,max} 

\newcommand{\RNum}[1]{\uppercase\expandafter{\romannumeral #1\relax}} 

\newcommand{\matr}[1]{\boldsymbol{#1}}

\newcommand{\msigma}{\matr{\sigma}} 
\newcommand{\tsigma}{\tilde{\matr{\sigma}}} 
\newcommand{\mA}{\matr{A}}
\newcommand{\mb}{\matr{b}}
\newcommand{\mC}{\matr{C}}
\newcommand{\mD}{\matr{D}}

\newcommand{\mI}{\matr{I}}
\newcommand{\mLam}{\matr{\Lambda}}

\newcommand{\mP}{\matr{P}} 
\newcommand{\mS}{\matr{S}}
\newcommand{\mSigma}{\matr{\Sigma}}

\newcommand{\mTheta}{\matr{\Theta}}
\newcommand{\mV}{\matr{V}}
\newcommand{\mx}{\matr{x}}

\newcommand{\mZ}{\matr{Z}}
\newcommand{\dgamma}{\dot{\gamma}}
\newcommand{\raumR}{\mathbb{R}}

\newcounter{bla}

\journal{Computer Physics Communications}

\begin{document}
\pagenumbering{arabic}

\begin{frontmatter}



\title{Molecular dynamics simulations in hybrid particle-continuum schemes: \\Pitfalls and caveats}


\author[a]{S.~Stalter}
\author[b]{L.~Yelash}
\author[c]{N.~Emamy}
\author[d]{A.~Statt}
\author[b]{M.~Hanke}
\author[b]{M.~Luk\'a\v{c}ov\'a-Medvid'ov\'a\corref{author1}}
\author[a]{P.~Virnau\corref{author2}}

\cortext[author1]{Corresponding author.\\\textit{E-mail address:} lukacova@mathematik.uni-mainz.de}
\cortext[author2] {Corresponding author.\\\textit{E-mail address:} virnau@uni-mainz.de}
\address[a]{Institute of Physics, Johannes Gutenberg University, Staudingerweg 9, 55128 Mainz, Germany}
\address[b]{Institute of Mathematics, Johannes Gutenberg University, Staudingerweg 9, 55128 Mainz,Germany}
\address[c]{Institute for Parallel and Distributed Systems, University of Stuttgart, Universit\"{a}tsstra\ss e 38, Stuttgart, Germany}
\address[d]{Department of Chemical and Biological Engineering, Princeton School of Engineering and Applied Science, Princeton, NJ 08544}

\begin{abstract}

Heterogeneous multiscale methods (HMM) combine molecular accuracy of particle-based simulations with the computational efficiency of continuum descriptions to model flow in soft matter liquids. In these schemes, molecular simulations typically pose a computational bottleneck, which we investigate in detail in this study. We find that it is preferable to simulate many small systems as opposed to a few large systems, and that a choice of a simple isokinetic thermostat is typically sufficient while thermostats such as Lowe-Andersen allow for simulations at elevated viscosity. We discuss suitable choices for time steps and finite-size effects which arise in the limit of very small simulation boxes. We also argue that if colloidal systems are considered as opposed to atomistic systems, the gap between microscopic and macroscopic simulations regarding time and length scales is significantly smaller. We propose a novel reduced-order technique for the coupling to the macroscopic solver, which allows us to approximate a non-linear stress-strain relation efficiently and thus further reduce computational effort of microscopic simulations.
\end{abstract}


\end{frontmatter}



\section{Introduction}
\begin{small}  
Modeling and computational simulation of soft matter liquids remains a challenging problem because these fluids may exhibit complex non-Newtonian effects, such as shear-thinning/thickening, viscoelasticity or flow-induced phase transition.
Such complex behavior is attributed to microstructure changes in fluids when a system is subject to an external mechanical shear force \cite{Malkin2013, Newstein_1999}. Therefore, computational modeling of soft-matter fluids has to necessarily take into account microscopic effects in order to obtain reliable numerical solutions.

Clearly, the most accurate description of soft-matter fluids can be obtained by the molecular dynamics (MD). However, such microscale description is
computationally inefficient, if large scale regions in space and time need to be simulated. To overcome this restriction and to obtain
practically tractable simulation techniques hybrid molecular-continuum methods have been proposed in the literature aiming
in combining the best attributes of both parts:  the molecular accuracy with the computational efficiency of continuum models.

Bridging the large range of dynamically coupled scales is a fundamental challenge that is a driving force in the development of new mathematical algorithms. In general, hybrid models  can be divided in two groups: based on the Eulerian-Lagrangian decomposition
or on domain decomposition. In the first type the Lagrangian-type particles are embedded in the Eulerian fluid description, see, e.g., \cite{emamy_2017, Ren_2005, Yasuda_2010}. The second type of the methods is based on the domain decomposition into a small accurate atomistic region embedded into a coarser macrosopic model, see, e.g., \cite{Fedosov_2009}. In the literature we can find several hybrid models combining particle dynamics with the macroscopic continuum model, see, e.g., the hybrid heterogeneous multiscale methods described in \cite{Weinan2007, E2003, E2004, E_2005, Ren_2007, Ren_2005, Yasuda_2010}, the triple-decker atomistic-mesoscopic-continuum method \cite{Fedosov_2009}, the seamless multiscale methods \cite{E2007,e_2007}, the equation-free multiscale methods \cite{Kevrekidis_2003,Kevrekidis_2009} or the internal-flow multiscale method \cite{lockerby2, lockerby1}. In \cite{koumoutsakos}  a overview of multiscale flow simulations using particles is presented. The essential question that arises in building
a coupled multiscale method is how micro- and macroscopic
models are linked together, i.e., how projection/lift (or compression/reconstruction) operators are defined and implemented.

On the artificial boundary of the particle domain embedded into the macroscopic domain following typical strategies of constraint dynamics can be found in the literature: the Maxwell buffer \cite{patera_1997}, the relaxation dynamics \cite{connel_thompson_1995}, the least constraint dynamics \cite{nie_2004}, and the flux imposition \cite{flekkoy_2000}.
Truncation of the microscopic domain is realized by imposing suitable boundary conditions.
Non-periodic boundary conditions involve particle insertions and
deletions, special wall reflections and body force terms, see  \cite{Fedosov_2009} for more details.
The deformation of the boxes mimics the time evolution of the control volume element in continuum and requires an adaption of standard Lees-Edwards periodic boundary conditions \cite{Todd1998}.

In order to extract mean flow field information
from particle-based simulations averaging needs to
be performed after a specified number of time steps. For example, the required rheological information for the stress tensor
is calculated using the Irving-Kirkwood expression \cite{Irving_1950} and passed to the macroscopic continuum model.

In fact all of these techniques can be considered as hybrid particle-continuum methods under the statistical influence of microscale effects since coefficients in coarse-grained equations are estimated from data that are obtained from microscale simulations. As demonstrated in
\cite{lockerby1} the sensitivity of the accuracy of a solution, as well as the computational speed-up over a full molecular simulation, is dependent on the degree of scale separation that exists in a problem. For the case when processes occurring on a small scale are only loosely coupled with the behavior on a much larger scale and the so-called {\sl scale separation} in the flow direction occurs, the hybrid multiscale schemes can be successfully applied, see \cite{lockerby2, lockerby1, Weinan2007, E2004, emamy_2017, Ren_2007, Ren_2005, yasuda2008model, Yasuda_2010, Yen_2007} and the references therein.

A complementary approach, the fluctuating hydrodynamics goes beyond the mean flow field of the hybrid simulation. In this case, the statistical influence of microscale effects is explicitly taken into account in the macroscopic flow equations leading to the stochastic partial differential models, such as the Landau-Lifshitz-Navier-Stokes system \cite{delgado_2007, donev_2010, convey_2007}. We refer reader to a recent study \cite{karniadakis_2016}, where the errors in the fluctuations, due to both the truncation of the domain and the constraint
dynamics performed on the artificial boundary are analysed for hybrid shear flow simulations.

In our recent paper \cite{emamy_2017} we have developed a novel hybrid multiscale method that is based on the combination of the discontinuous Galerkin (dG) method and molecular dynamics (MD) in order to simulate complex fluids, such as colloids in a Newtonian solvent.
It has been shown that the method can be applied successfully to complex fluids when scale separation occurs and we can assume that
the statistical influence of the microscale can be controlled on the macroscale.  Our dG-MD hybrid method combines the following advantages (i) for macroscopic flow equations the dG method is applied which allows more
flexible discretization including per-cell momentum conservation, (ii) the reduced order techniques are included in order to control the number of needed but computationally expensive MD simulations. The main goal of  the present paper is to focus
on the molecular dynamics part, which typically poses the bottleneck in these hybrid molecular-continuum approaches. We will discuss strategies, which minimize the computational effort in the particle-based simulations and discuss optimum choices for thermostats, time steps and relate time and length scales from simulations to experiments. Moreover we investigate the coupling of the microscopic simulation data to the macroscopic flow solver and propose a novel reduced-order strategy based on the
combination of the proper orthogonal decomposition, the regularized least-square approximation and a suitable greedy algorithm
to approximate the unknown nonlinear stress-strain function efficiently.
This is the first time that the reduced-order technique is used in the context of hybrid simulation methods.
As a test case we investigate Couette and Poiseuille flow in two and three dimensions.

\section{Microscale (particle-based) simulations}
\label{sec:md}

In non-equilibrium Molecular Dynamics (MD) \cite{Frenkel}, we simulate colloidal particles with a coarse-grained model. Since we simulate a one dimensional flow, standard Lees-Edwards periodic boundary conditions \cite{lees} are applied. In our case, we shear in $z$-direction, and apply the velocity-Verlet algorithm to solve the equations of motion \cite{Swope:1982}.

\begin{figure}[h] 
	\centering
	\includegraphics[width=0.3\textwidth]{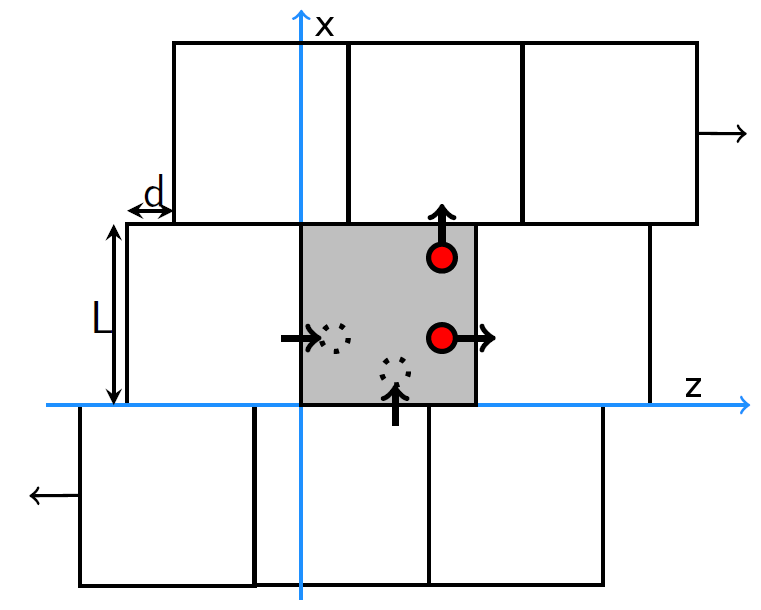}
	\caption{Lees-Edwards boundary conditions. In the simulations, we shear in $z$ direction.}
\end{figure}

The stress-tensor is calculated via the Irving-Kirkwood formula \cite{Irving_1950}, using the peculiar velocities of particles $\vec{v}_i=\tilde{\vec{v}}_i-\vec{v}_{i,S}$, where $=\tilde{\vec{v}}_i$ is the total and $\vec{v}_{i,S}$ the streaming velocity of particle $i$, respectively.
\begin{equation}
\sigma_{\alpha \beta}=-\frac{1}{V}\left(\sum _i^N\left(m_iv_{i,\alpha}v_{i,\beta}\right)+\sum _i ^N \sum _{j>i}^N \left(r_{ij,\alpha}F_{ij,\beta}\right)\right)~,
\end{equation}
where $\vec{r}_{ij}$ and $F_{ij}$ are the distance and the force between particle $i$ and $j$. The pressure corresponds to
\begin{equation}
p=-\frac{\text{Tr}\mathbf{\sigma}}{3}~.
\end{equation}
The dynamic viscosity is
\begin{equation}
\eta =\left|\frac{\sigma _{13}}{\dot{\gamma}}\right|~,
\end{equation}

\noindent 
where $\dot{\gamma}$ is the shear rate.
Note that $\sigma _{13}$ will be used later on to couple microscopic simulations to the macroscopic solver.

Colloids are treated as hard spheres. The interaction of two particles is simulated with a Weeks-Chandler-Andersen potential (WCA), which corresponds to the repulsive part of the Lennard-Jones potential:
\begin{equation}
	V_{WCA}(r)=4\varepsilon \left[\left(\frac{\sigma}{r}\right)^{12}-\left(\frac{\sigma}{r}\right)^6+\frac{1}{4}\right]~, \quad r<r_C=\sqrt[6]{2}\sigma~.\label{eq:wca}
\end{equation}
Two particles in the interaction radius $r_C$ reject each other. Outside this radius $r_C$, the potential is $0$. In formula \eqref{eq:wca} $\varepsilon$ corresponds to the well depth of the Lennard-Jones potential. In our simulations, we set $\varepsilon=1k_BT$. $\sigma$ is the typical diameter of a colloid and is used as the length scale of the MD simulation.


It is worth noting that if particles represent colloids, $\sigma \approx 1~\mu m$ and typical time scales $t_{MD}$ are in the order of $10^{-4}~s$, which is close to experimental relaxation times. If particles represent atoms instead ($\sigma \approx 10^{-10}~m$), the gap between time scales of the simulation and experimental (macroscopic) time scale spans many orders of magnitude ($t _{MD}\approx 10^{-12}s$). See Appendix~\ref{app:sec:Mapp}.

Shearing a system leads to friction, which heats up the sheared system. Since we want to simulate in the NVT-ensemble, we need to cool the system down to an initial temperature $T_0$. In the following we discuss two thermostats.
The Lowe-Andersen thermostat (LAT) \cite{lowe} conserves momentum and is Galilean invariant. In this thermostat, after solving the equations of motion via the velocity-Verlet algorithm, particles undergo a "bath" collision with a probability of $\Gamma \Delta t$ (with $0<\Gamma \Delta t \leq 1$).

Two particles within an interaction range $r_{ij}<r_C$ have the possibility to get a "kick" along the center of mass, where the "kick" is taken from a Maxwell-Boltzmann distribution.

\begin{equation}
	\vec{v}_{ij}^\prime = \xi _{ij} \sqrt{2k_BT} \hat{r}_{ij}
\end{equation}
where $\xi _{ij}$ is a random number with unit variance and $\hat{r}_{ij}$ the normalized particle distance. The new velocities for particles $i$ and $j$ after a bath collision are:
\begin{equation}
	\vec{v}_i^\prime = \vec{v}_i+\vec{\Delta}_{ij}
\end{equation}
and
\begin{equation}
\vec{v}_j^\prime = \vec{v}_j-\vec{\Delta}_{ij},
\end{equation}
where
\begin{equation}
	2\vec{\Delta}_{ij} = \hat{r}_{ij} \left(\vec{v}_{ij}^\prime-\vec{v}_{ij}\right)\cdot \hat{r}_{ij}
\end{equation}
which leads to momentum conservation. In the following we set the bath collision probability to $\Gamma \cdot \Delta t=0.001$.

The second thermostat we would like to discuss is the simple isokinetic (ISO) thermostat \cite{Yasuda_2010}, a popular choice amongst hybrid simulation schemes. After each velocity-Verlet integration step, we compute the temperature of a system of $N$ particles:

\begin{equation}
\frac{3}{2}Nk_BT=\frac{1}{2}\sum _i^N m_i v_i^2
\end{equation}
and rescale the velocities with the factor
\begin{equation}
	\lambda = \sqrt{\frac{T_0}{T}}
\end{equation}
where $T_0$ is the temperature at which we want to simulate. (Here, $T_0=1\frac{\varepsilon}{k_B}$.)

Additionally, we implement the SLLOD algorithm \cite{Yasuda_2010}, which applies the shear profile to the equations of motion. Like this, we reach the steady state even faster. The equations of motion change to
\begin{equation}
	\dot{\vec{q}}=\vec{p}+\bigtriangledown \vec{u}\cdot\vec{q}\label{eq:q}
\end{equation}
\begin{equation}
	\dot{\vec{p}}=\frac{\vec{f}}{m}-\bigtriangledown \vec{u}\cdot\vec{p}\label{eq:p}
\end{equation}
where $\bigtriangledown \vec{u} $ is the matrix representation of the applied stress:
\begin{equation}
	\bigtriangledown \vec{u} = \left(\begin{array}{ccc}
	0 & 0 & 0\\
	0 & 0 & 0\\
	\dot{\gamma} & 0 & 0
	\end{array}\right)
\end{equation}

since we shear the $xy$-plane in the $z$-direction.

\begin{figure}[hbt] 
	\centering
	\hspace{-0.75cm}\subfigure[Lowe-Andersen]{\includegraphics[width=0.25\textwidth, clip]{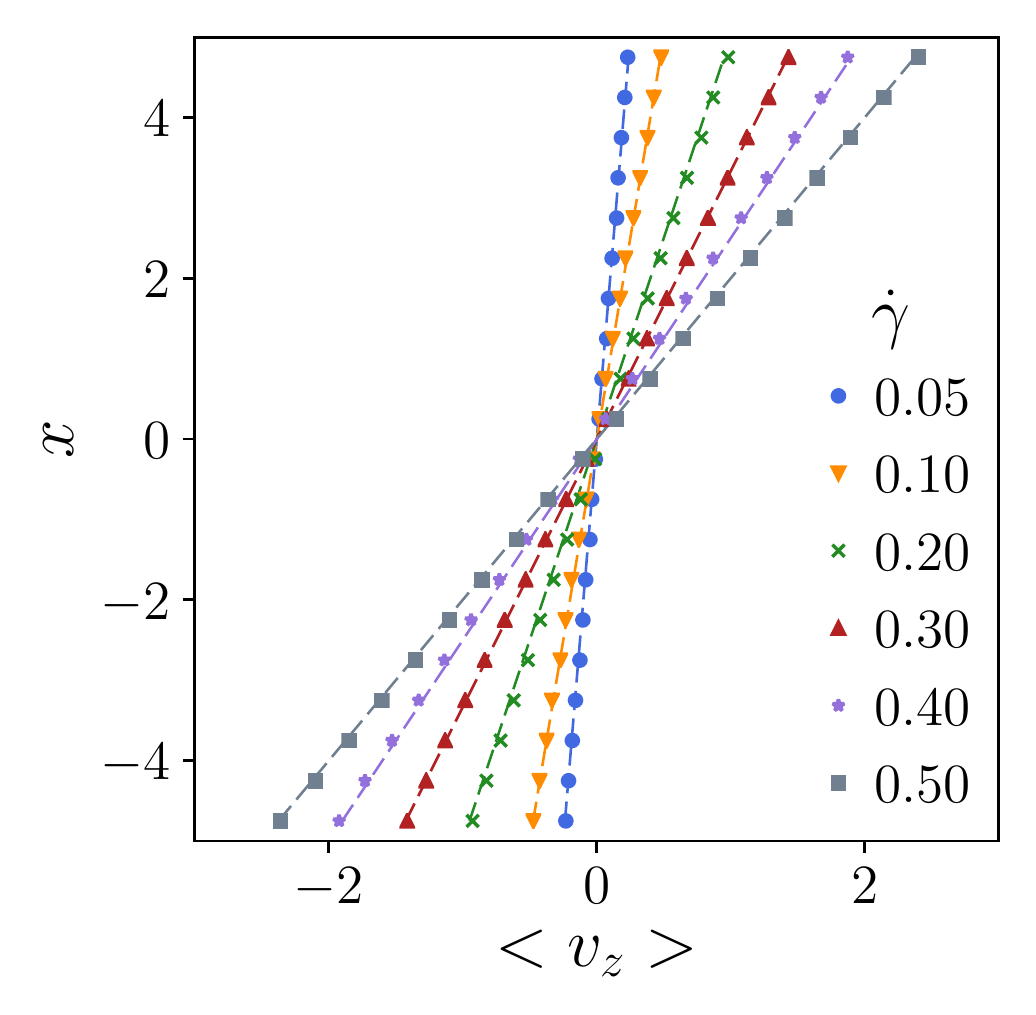}}
	\subfigure[Isokinetic with SLLOD]{\includegraphics[width=0.25\textwidth, clip]{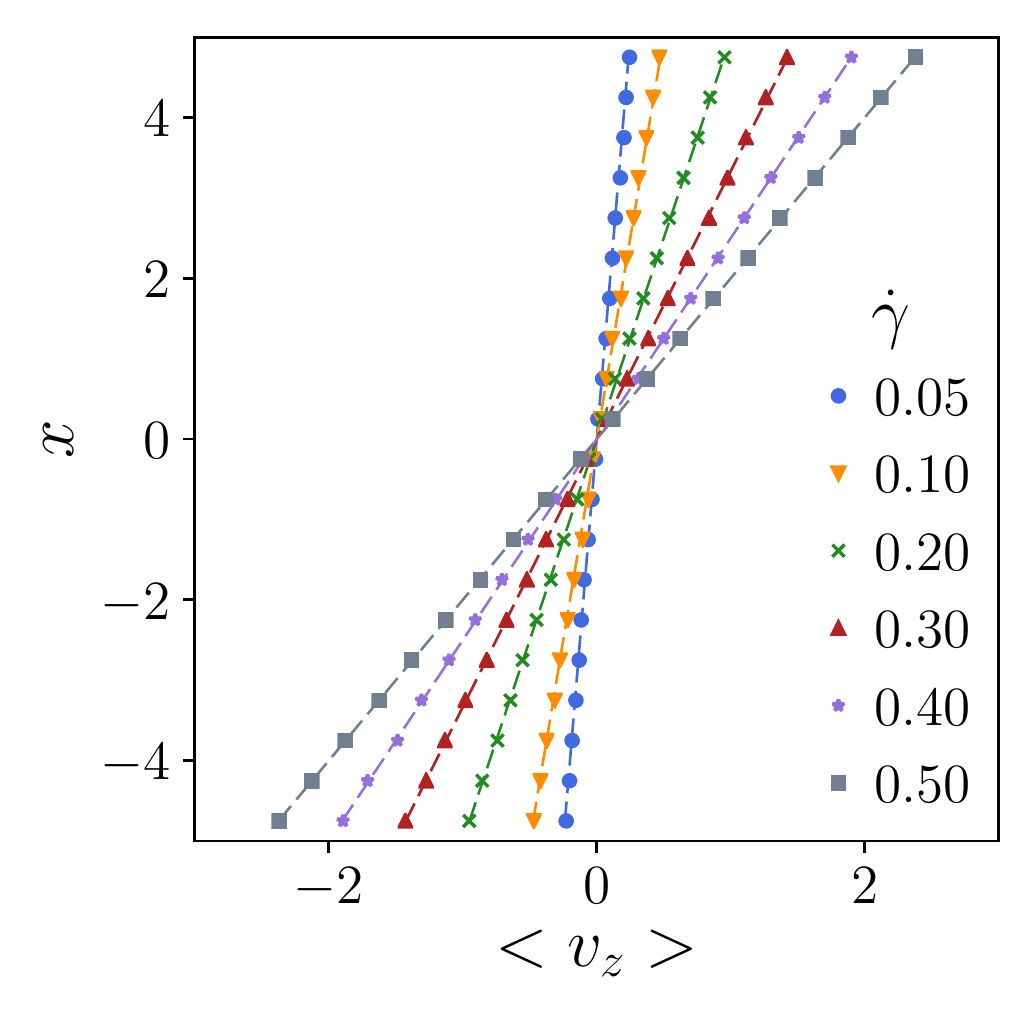}}
	\subfigure[Time evolution of the $\sigma _{13}$ component.]
	{\includegraphics[width=0.4\textwidth]{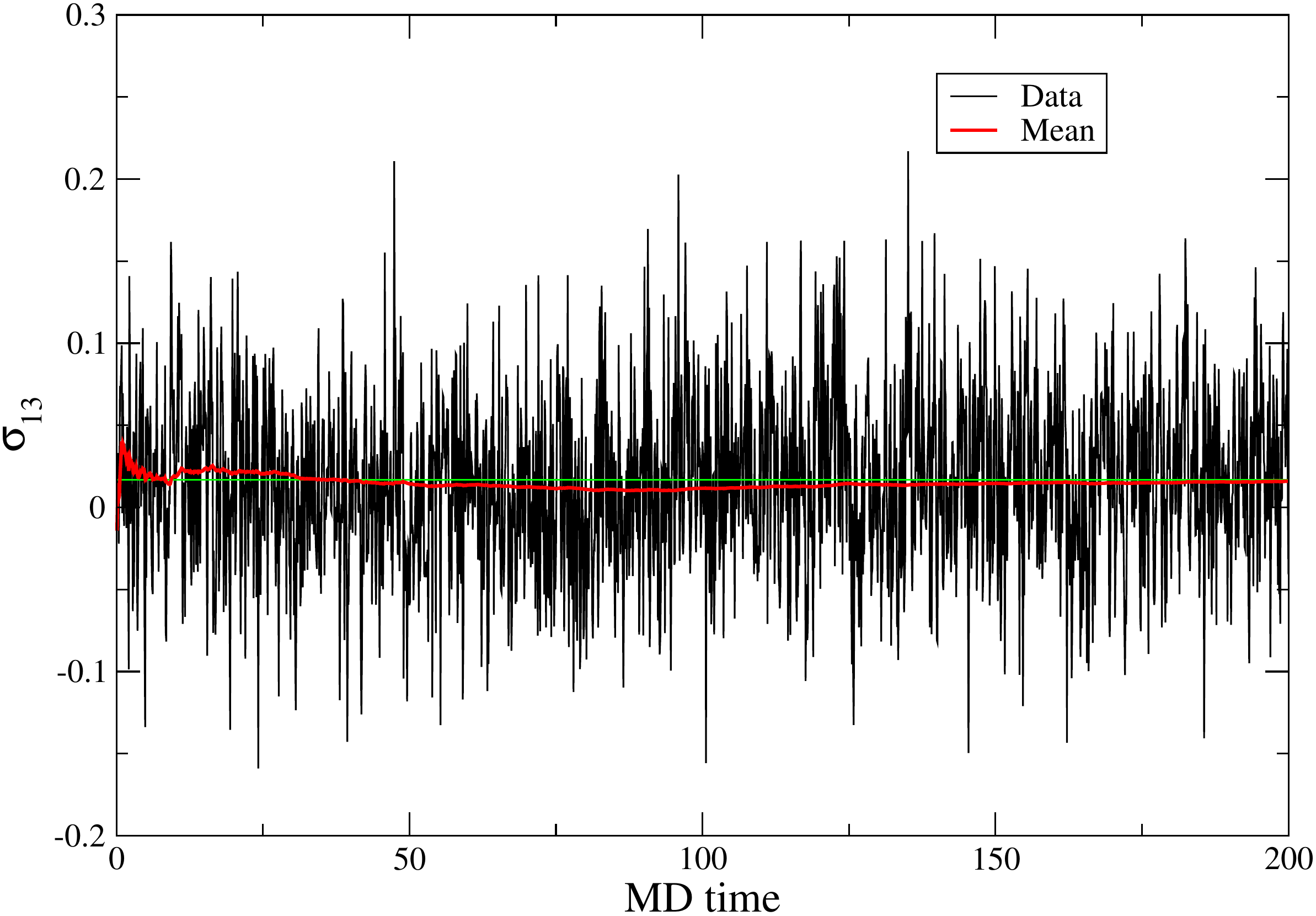}}
	\caption{Shear profiles for a system at particle density $\rho = 0.4$ (containing $400$ particles) and varying shear rates. The integration time step was chosen to be $10^{-4}$, after a relaxation time of $200$ MD-times, the shear profiles are taken for $1\,500$ MD-times. (a) Lowe-Andersen thermostat with a bath collision probability of $\Gamma\cdot\Delta t=0.001$. (b) Isokinetic thermostat with SLLOD. {Symbols: MD results, lines: linear profiles expected according to $\langle v_z\rangle=\dot{\gamma}x$.} In (c) the time to steady state for the $\sigma_{13}$ component at a shear rate of $\dot{\gamma} =0.05$ is shown. 
		The instantaneous data of the $\sigma_{13}$ component is black, the mean at a certain MD-time is red, while the overall mean is colored in green.}
	\label{fig:profiles}
\end{figure}

Here, we extend the equations of motion for the half step, e.g., one equation for the generalized coordinates $\vec{q}^{n+\frac{1}{2}}$ due to the new terms of the SLLOD method in \eqref{eq:q} and \eqref{eq:p}. These equations will then be needed for the full step. For a given $\Delta \vec{u}$, the half steps look like:
\begin{align}
	\vec{q}^{n+\frac{1}{2}}&=\vec{q}^n+\frac{1}{2}\Delta t \left( \vec{p}^n+\dot{\gamma}~\left( \vec{q}_x^n \cdot \hat{e}_z\right)\right)\\
	\vec{p}^{n+\frac{1}{2}}&=\vec{p}^n+\frac{1}{2}\Delta t \left( \vec{F}^n -\dot{\gamma}~\left(\vec{p}_x^n\cdot \hat{e}_z\right)\right)
\end{align}
where $n$ is the current time step, $n+\frac{1}{2}$ is the time of the half step and $n+1$ is the full step. The index $x$ in $\vec{q}$ and $\vec{p}$ correspond to the $x$ coordinate of the vector.

Analogously to the half step, we have to modify the equations of motion for $\vec{q}$, which includes the half step $\vec{q}^{n+\frac{1}{2}}$, consistent with the time level of $\vec{p}^{n+\frac{1}{2}}$:

\begin{align}
	\vec{q}^{n+1}&= \vec{q}^n+\Delta t\left(\vec{p}^{n+\frac{1}{2}} + \dot{\gamma}~\left(\vec{q}^{n+\frac{1}{2}}_x \cdot \hat{e}_z \right)\right)~,\\
	\vec{F}^{n+1}&= -\vec{\nabla} V(\lvert \vec{q}^{n+1}\rvert) ~,	\\
	\vec{p}^{n+1}&= \vec{p}^{n+\frac{1}{2}} + \frac{1}{2} \Delta t \left( \vec{F}^{n+1} - \dot{\gamma} \left(\vec{p}^{n+1} _x \cdot \hat{e}_z\right)  \right).
\end{align}

This method takes advantage of the Crank-Nicolson method. As pointed out in \cite{Leimkuhler2013} it allows larger time steps and yields one order smaller errors.

Figure~\ref{fig:profiles} shows averaged shear profiles obtained with (a) the Lowe-Andersen thermostat and (b) the isokinetic thermostat with SLLOD in agreement with the shear rate imposed by Lees-Ewards boundary and SLLOD conditions. Note that SLLOD conditions in the isokinetic case impose a linear shear profile, which takes somewhat longer to emerge for the Lowe-Andersen thermostat, and increases further for lower shear rates. Figure~\ref{fig:profiles}c displays a typical relaxation of the off-diagonal component of the stress tensor $\sigma _{13}$ after turning on shear. The component relaxes in less than 200 MD times as indicated by the running mean (red curve). We have verified that this holds for various shear rates and densities for the small system sizes considered in this study, and in all simulations discussed from now on, measurements start after 200 MD times to omit influences of the relaxation process.

In Figure~\ref{fig:ThermoQuantities} we check to which extent the Lowe-Andersen thermostat is able to thermalize the system for our given set of simulation parameters ($\Gamma \cdot \Delta t=0.001$). For high densities and shear rates, deviations in the temperature (a), trace of the stress tensor (b) and viscosity (c) are noticeable in comparison to the isokinetic thermostat, which strictly enforces temperature. As already stated in the original publication on the Lowe-Andersen thermostat \cite{lowe}, the viscosity of such a system is somewhat elevated (Figure~\ref{fig:ThermoQuantities}d), and the surplus viscosity is only expected to vanish in the limit of small $\Gamma$ and large time steps, which again counteracts efforts to keep temperature and pressure fixed. On the other hand, this enables exploration of fluids with somewhat larger viscosities in the context of hybrid-continuum schemes.

\begin{figure*}[hbtp] 
	\centering
	\subfigure[Temperature vs. shear rate.]{\includegraphics[width=0.45\textwidth]{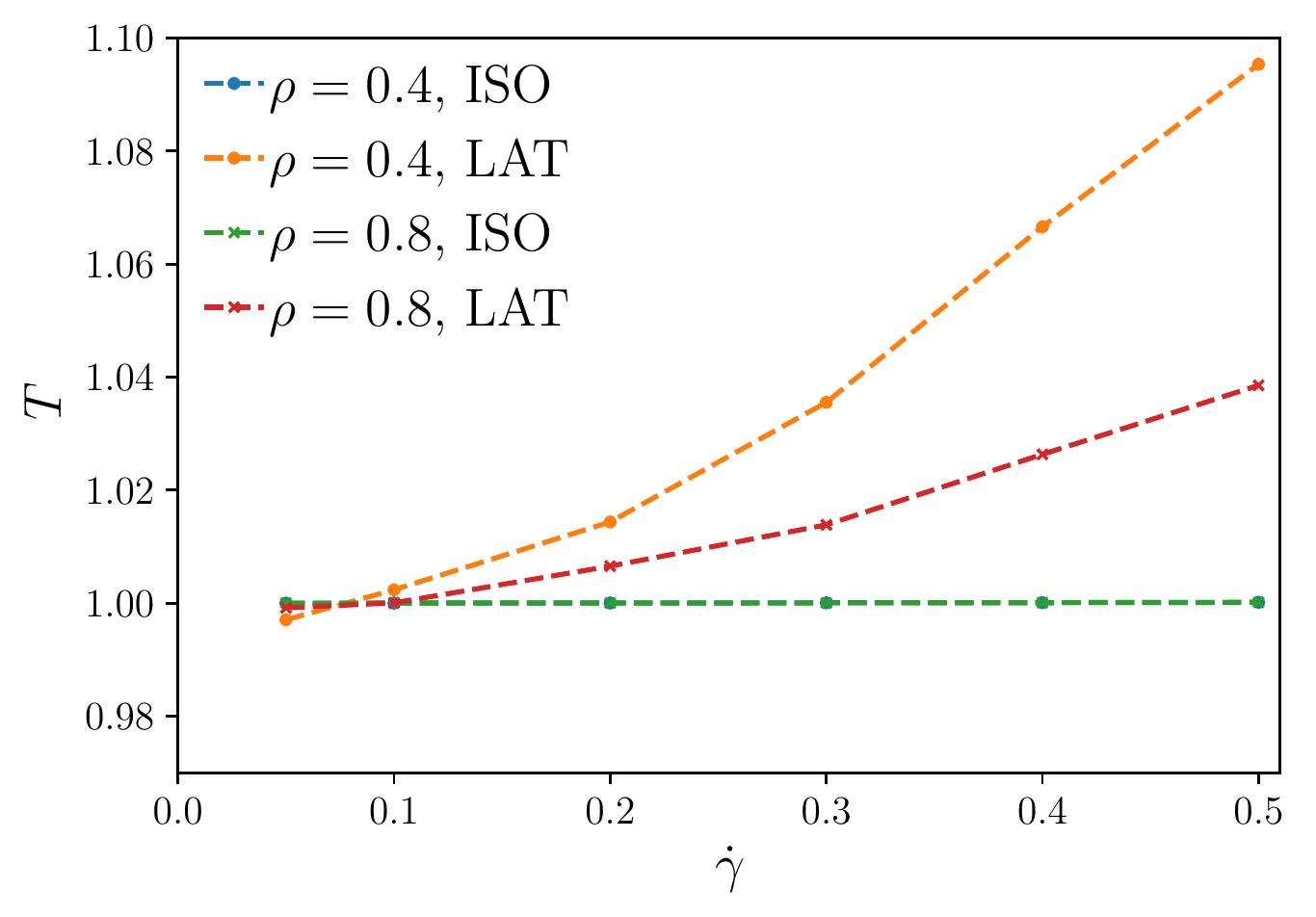}}
	\subfigure[Pressure vs. shear rate.]{\includegraphics[width=0.45\textwidth]{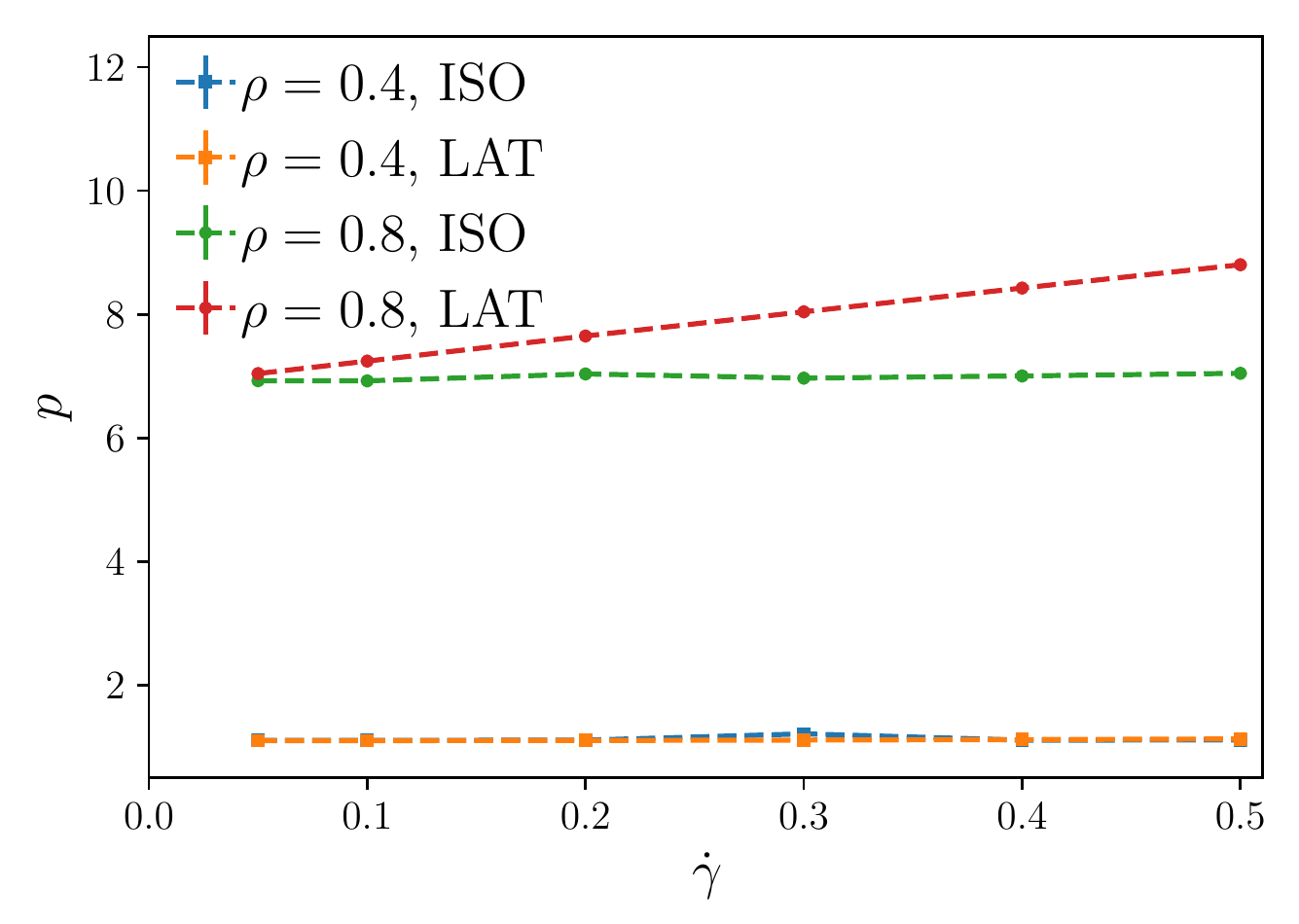}}
	\subfigure[Viscosity vs. shear rate.]{\includegraphics[width=0.45\textwidth]{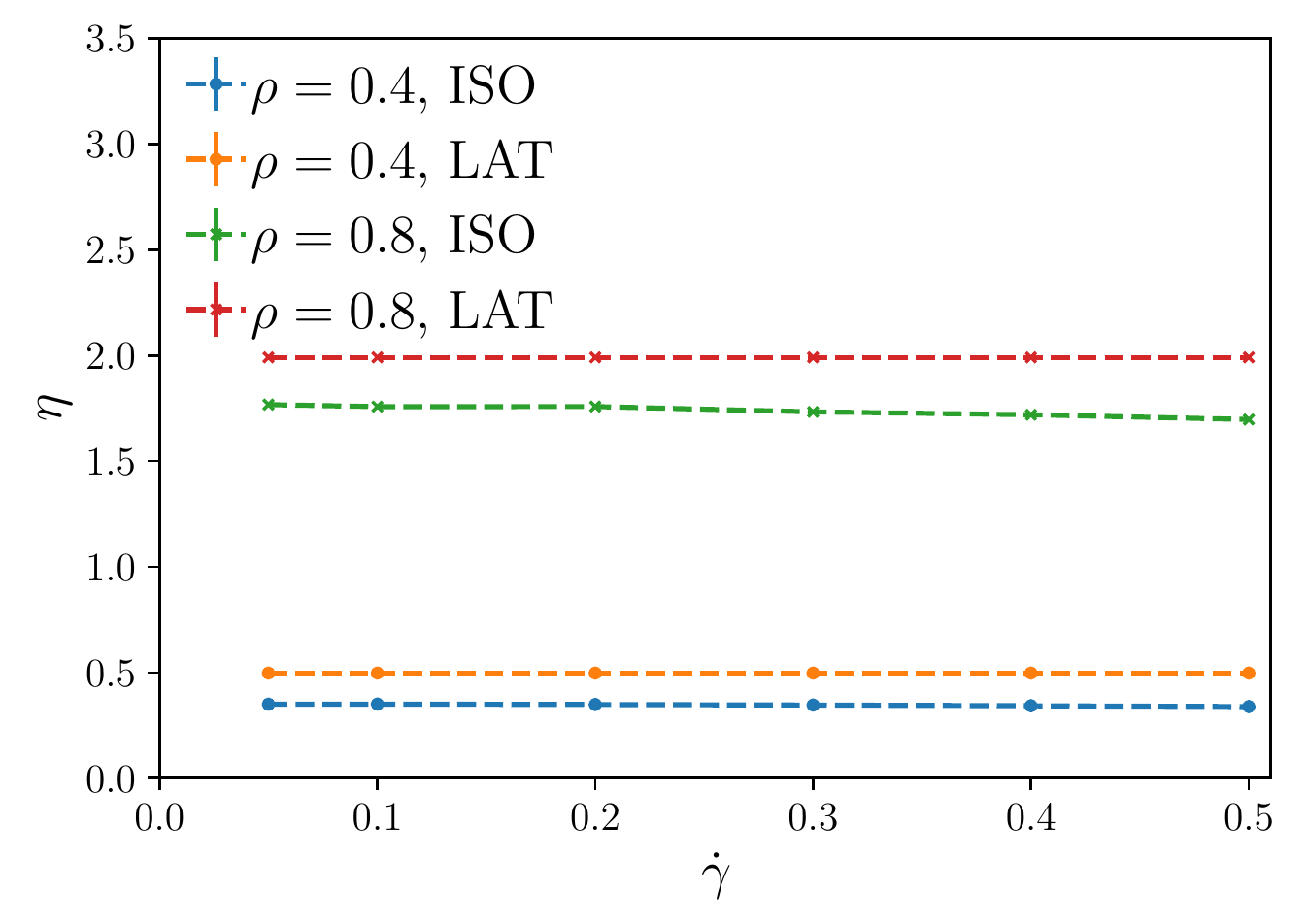}}
	\subfigure[Viscosity vs. $\Gamma$.]{\includegraphics[width=0.45\textwidth]{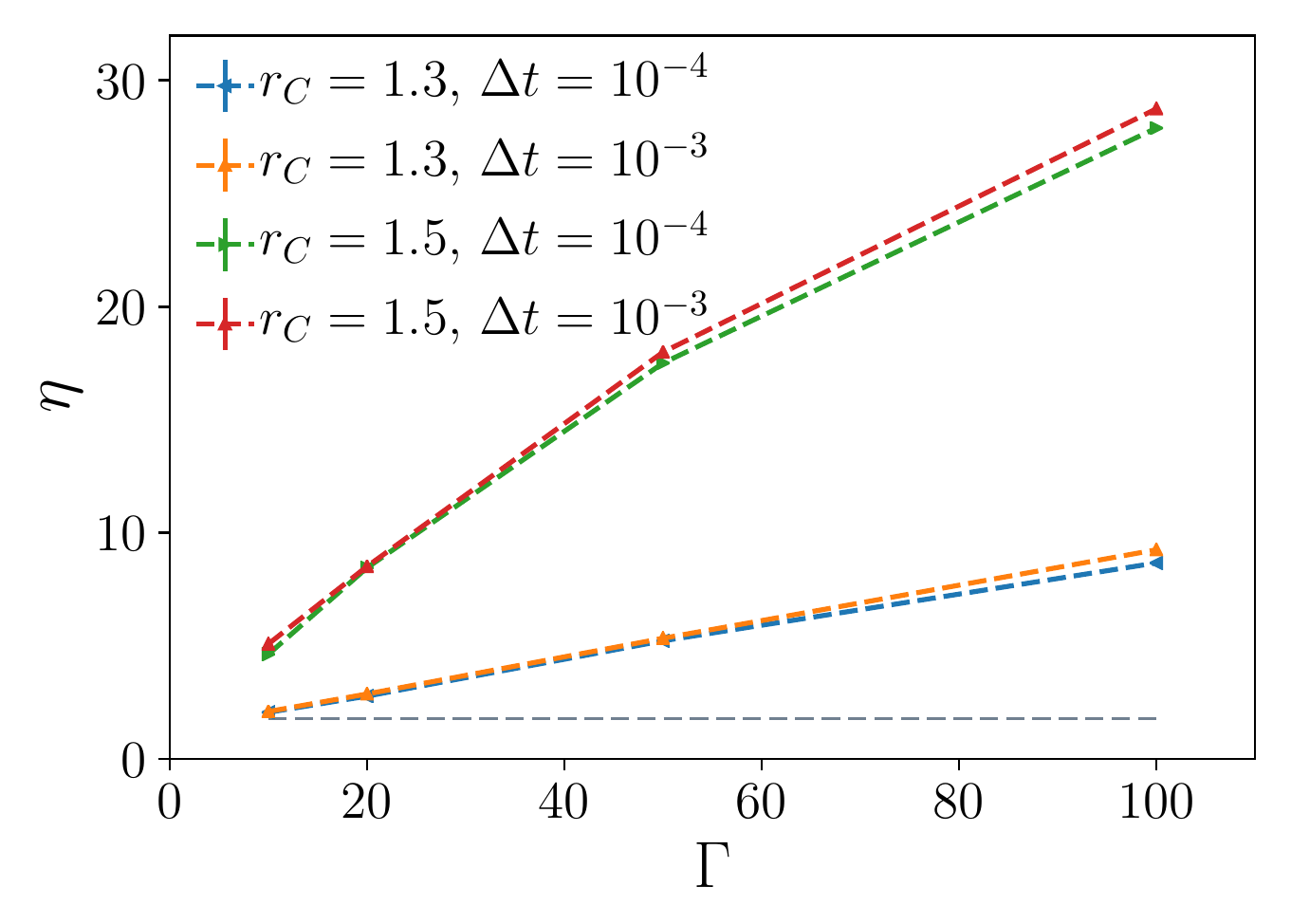}}
	\caption{Comparison of Lowe-Andersen and isokinetic thermostat with SLLOD. (a) Temperature, (b) pressure and (c) viscosity as a function of shear rate after relaxation (of $200$ MD times). The system size is $10 \, \sigma$ in all directions. The time step is set to $10^{-4}$ for isokinetic and Lowe-Andersen thermostat. In the LAT, $\Gamma \cdot \Delta t$ is set to $0.001$. In d) the viscosity is shown as a function of bath collision frequency $\Gamma$ for different thermostat interaction radii at two different time steps $\Delta t = 10^{-4}$ and $\Delta t = 10^{-3}$. The grey dashed line corresponds to the viscosity obtained from simulations using the isokinetic thermostat and serves as a guide to the eye.}
	\label{fig:ThermoQuantities}
\end{figure*}

\begin{figure*}[hbtp] 
	\centering
	\subfigure[Trace of the stress vs. shear rate for different timesteps.]{\includegraphics[width=0.45\textwidth]
		{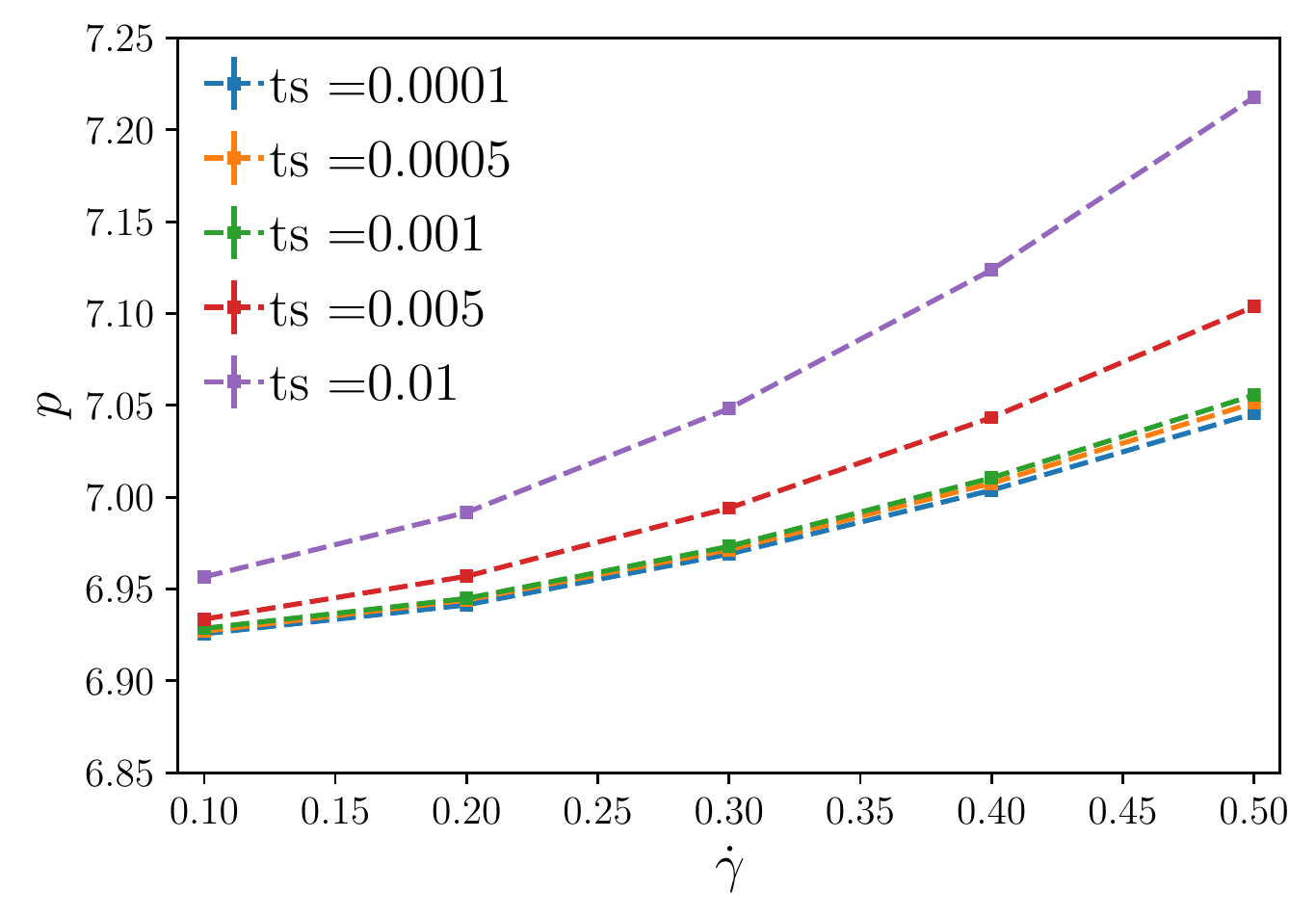}}
	\subfigure[Viscosity vs. shear rate for different timesteps.]{\includegraphics[width=0.45\textwidth]
		{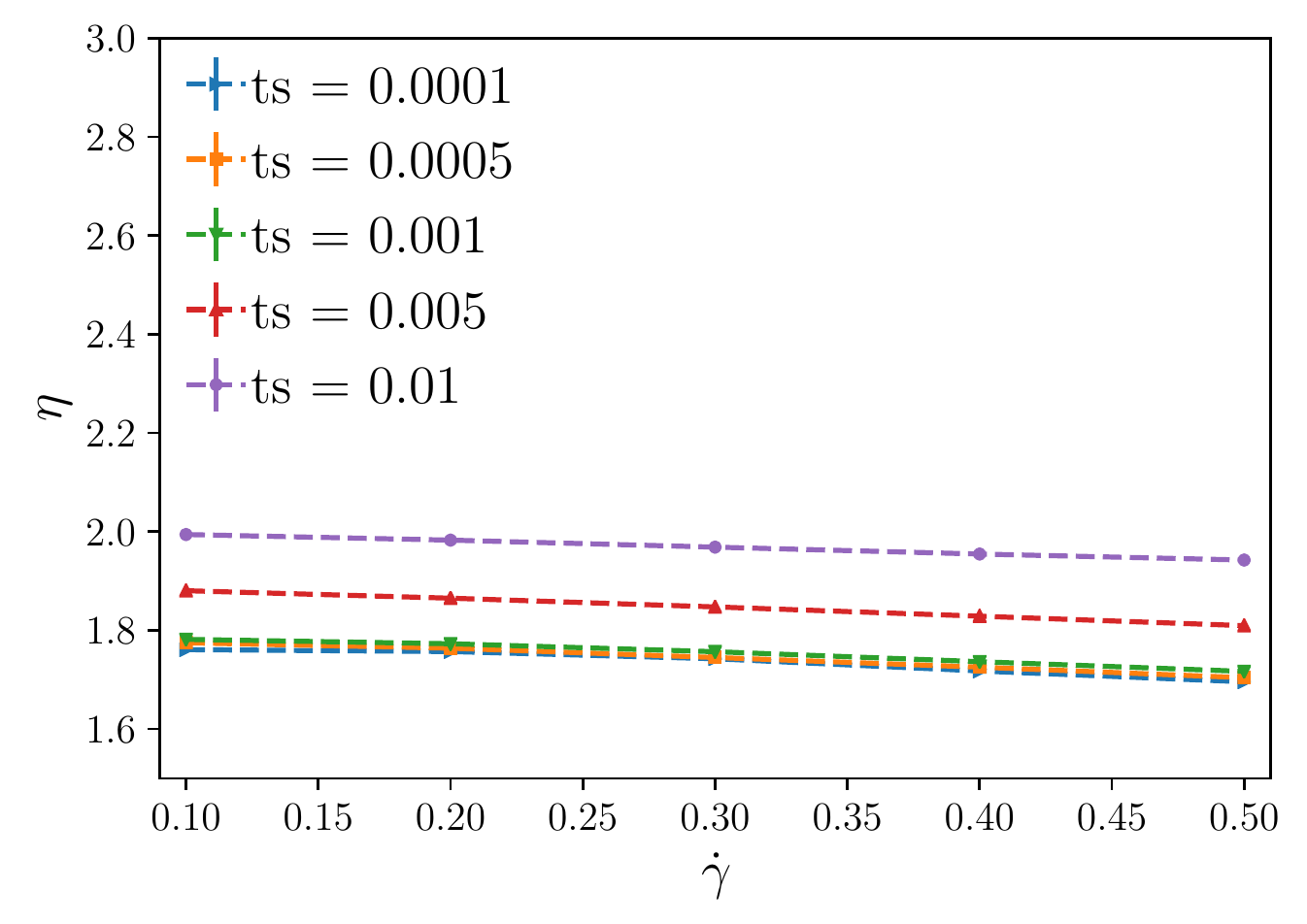}}
	\subfigure[Pressure vs. box length.]{\includegraphics[width=0.45\textwidth]{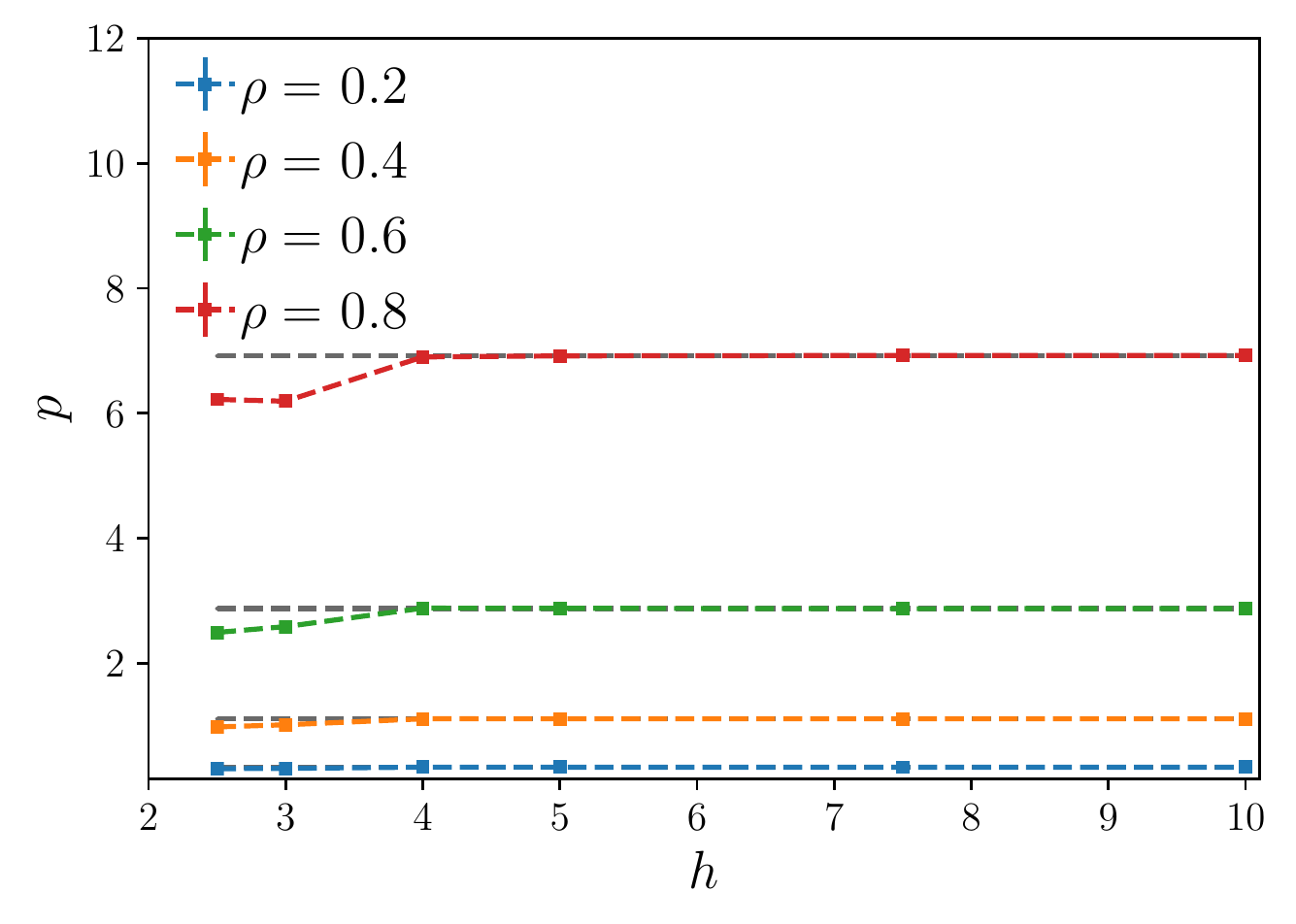}}
	\subfigure[Viscosity vs. box length.]{\includegraphics[width=0.45\textwidth]{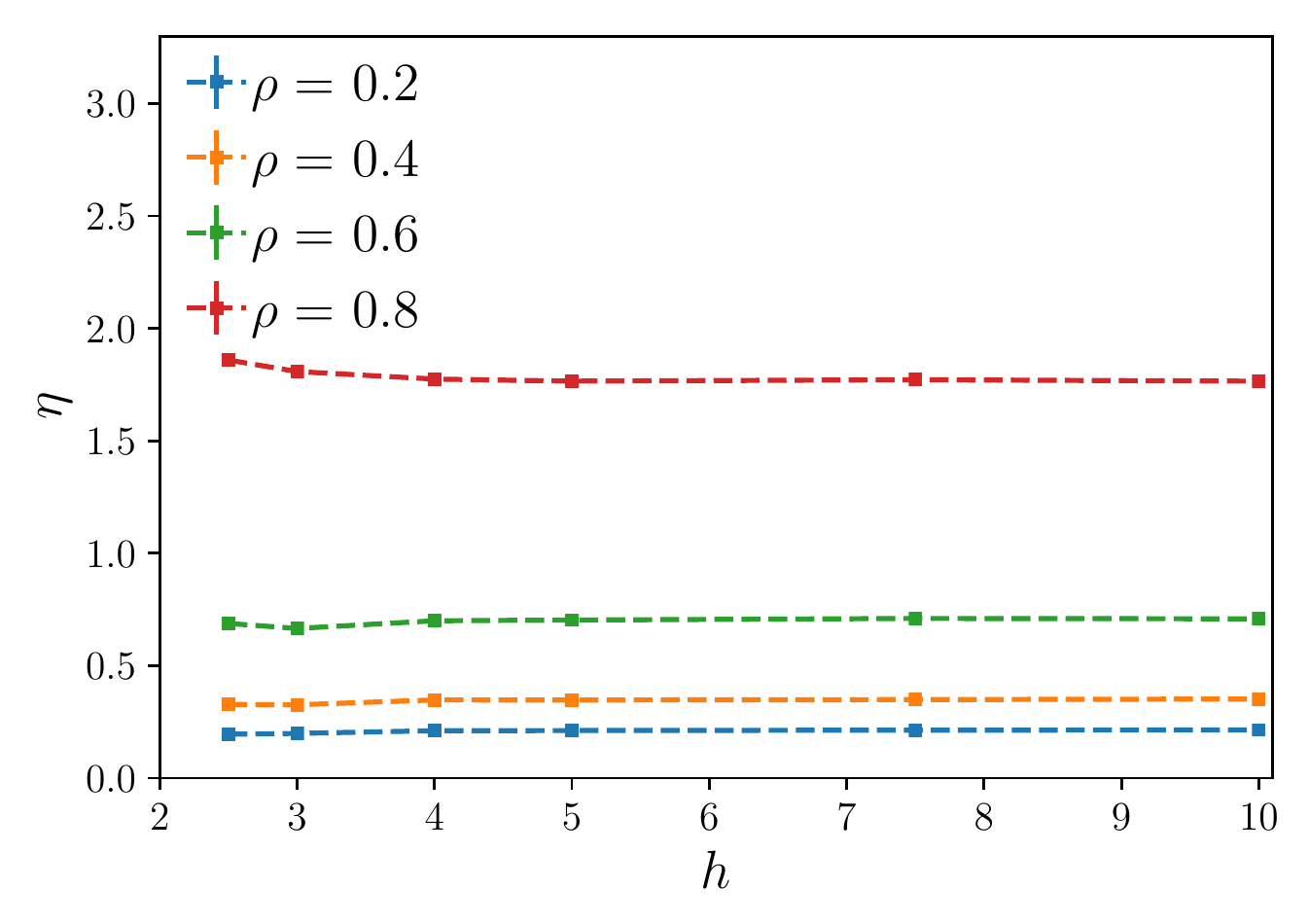}}
	\caption{Comparison of (a) trace of the stress, (b) viscosity as a function of shear rate after relaxation for different time steps. The density is $\rho=0.8$ in a box of dimensions $10\times 7.5\times 10\,\sigma ^3$, and comparison of (c) trace of the stress and (d) viscosity as a function of box length after relaxation with a shear rate of $\dot{\gamma}=0.1$. For different densities $\rho = 0.2, 0.4, 0.6, 0.8$ the $y$-dimension of the box is varied, while the other two dimensions are kept constant at $15\times 15 \,\sigma ^2$. The grey dotted line are the values for an unsheared system with box dimension $15\times 10\times 15 ~\sigma ^3$.}
	\label{fig:timeStep}
\end{figure*}

In such a scenario, the model needs to be considered in conjunction with the thermostat and thermostat parameters. (Note that even though viscosity is elevated, the system still behaves like a Newtonian fluid, i.e., the viscosity changes only little as a function of the shear rate (Figure~\ref{fig:ThermoQuantities}c).) If we are, however, interested in the viscosity of the actual model system or related properties, it is preferable to apply a thermostat, which does not enhance the latter (such as Lowe-Andersen or DPD). Even though the isokinetic thermostat with SLLOD (which we will use from now on) is somewhat unphysical, we are not interested in realistic dynamics, but properties of the steady state.

In attempt to minimize the computational effort, we investigate the dependence of the time step on pressure (Figure~\ref{fig:timeStep}a) and viscosity (Figure~\ref{fig:timeStep}b). While time steps larger than $10^{-3}$ (for the isokinetic thermostat) yield noticeable deviations in pressure and viscosity for dense systems and high shear rates, differences for a time step of $10^{-3}$ are probably acceptable in the context of a hybrid scheme. In this paper we use $10^{-4}$ if not noted differently.

When performing microscopic simulations in the context of a hybrid scheme, the question arises whether it is better to simulate a single large system or multiple systems of smaller size to gather statistics. From a computational point of view, the latter is typically to be preferred as MD simulations in practice often do not scale perfectly linear with the number of particles. Therefore, computational resources required to compute the contribution of ,e.g., a single particle to the stress tensor (over a given simulation period) are larger in a single large system when compared to multiple small ones. In addition, larger systems also require longer relaxation time. In Figure~\ref{fig:timeStep}c and d, we test the lower limits of system sizes for our model by decreasing the size of the simulation box in the direction perpendicular to the applied shear. As seen in Figure~\ref{fig:timeStep}c and d, finite-size effects start to play a role if the height of the box is smaller than four $\sigma$, which corresponds to four particle diameters. Of course, this value will change if the potential is longer-ranged or if particles become correlated, e.g., in the vicinity of a critical point.

\begin{figure}[hbt] 
	\centering
	\includegraphics[width=0.45\textwidth]{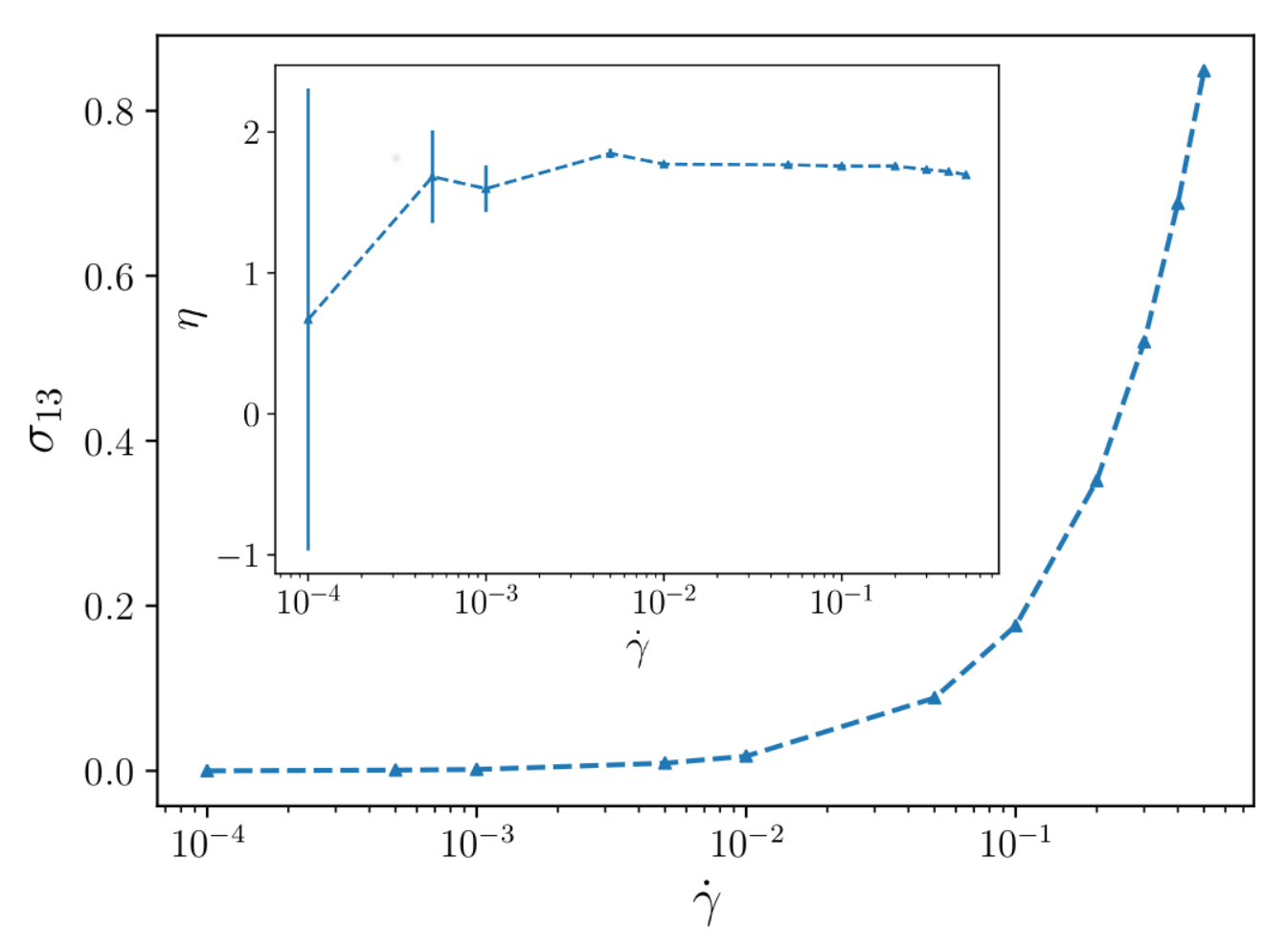}
	\caption{Viscosity $\eta$ (inset) and the $\sigma _{13}$ component of the stress tensor as a function of shear rate. Density is $\rho=0.8$ in a box of dimensions $10\times 7.5\times 10\,\sigma ^3$.}
	\label{fig:error}
\end{figure}

Finally, Figure~\ref{fig:error} shows the viscosity (inset) and the off-diagonal component of the stress-tensor $\sigma _{13}$ from which the viscosity is derived as a function of shear rate. As already indicated for large shear rates (Figure~\ref{fig:ThermoQuantities}), the viscosity changes only little over many orders of magnitude and behaves like a Newtonian fluid. It is worth noting, however, that the computational effort to obtain meaningful values for the viscosity increases manifold when the shear rate is reduced. While fluctuations in $\sigma _{13}$ are comparable across shear rates (not shown), fluctuations in the viscosity $\eta=\sigma _{13}/\dot{\gamma}$ increase by an order of magnitude if the shear is reduced accordingly. This increase in fluctuations translates into an increase of computational effort by two orders of magnitude if we want to keep errors at the same level.

In Section~\ref{sec:governingEquations}, we will briefly outline the macroscopic simulation. In Section~\ref{sec:hmm} we explain in detail the coupling of the micro and the macro level for which we need the data from Figure~\ref{fig:error} as input.

\section{Macroscale simulations}

 At the macroscopic level, the motion of the incompressible fluid flow is governed by the continuity and momentum equations

\label{sec:governingEquations}
   	\begin{subequations}
  		\begin{alignat}{3}
	  		\label{eq:Continuity}
  			&\nabla \cdot \vec{u}=0,&\quad &\textrm{in} \; \Omega \times [0, t_{\textrm{F}}]\\
  			\label{eq:Momentum}
  			&\rho\left(\frac{\partial \vec{u}}{\partial t}+ \vec{u}\cdot\nabla\vec{u}\right)=\frac{1}{Re} \nabla \cdot
  			\grvec{\sigma}+\vec{g},\quad& &\textrm{in} \; \Omega \times [0, t_{\textrm{F}}]\\
  			&\vec{u}=\vec{u}_D, & &\textrm{on} \quad \partial\Omega_D\\
  			&\vec{u}(t=0)=\vec{u}^{(0)} & &\textrm{in} \; \Omega,
  		\end{alignat}
  	\end{subequations}

\noindent
where $\vec{u}$ is the velocity vector, $\grvec{\sigma}$ the Cauchy stress tensor, $\vec{g}$ an external body force, $\rho$ the density, which is constant, and $Re$ the Reynolds number.
The computational domain $\Omega$ is surrounded by the boundary $\partial \Omega=\partial\Omega_D \cup \partial\Omega_P$, where the Dirichlet and periodic boundaries are considered, respectively.
In case of the Navier-Stokes equations for Newtonian fluids, $\grvec{\sigma}=-p \textbf{I}+\grvec{\tau}$,
where $\grvec{\tau}=\mu (\nabla \vec{u}+\nabla \vec{u}^T)$,
the above momentum equations reduce to
  	\begin{equation}
  		\label{eq:Navier-Stokes}
  		\rho\left(\frac{\partial \vec{u}}{\partial t}+ \vec{u}\cdot\nabla\vec{u}\right)=-\nabla p+\frac{1}{Re}(\mu\Delta\vec{u})+\vec{g} \quad \textrm{in} \; \Omega.  	
  	\end{equation}

\label{sec:explicit}
For the time integration of the continuity equation \eqref{eq:Continuity} and momentum equations \eqref{eq:Momentum} we apply the following multi-step projection method \cite{Emamy2017285}. Using the first-order Euler method we have

\begin{subequations}
\begin{alignat}{5}
\label{NonlinearStep-Euler}
&\text{\Romannum{1}} \quad &&\tilde{\vec{u}}=\Delta t \big(-\vec{u}^{(n)}\cdot\nabla\vec{u}^{(n)}+\frac{1}{Re} \nabla \cdot (\grvec{\sigma}^{(n)}/\rho)+\vec{g}^{(n)}/\rho\big),\quad &&\textrm{in} \; \Omega\\[1em]
\label{Poisson-Euler}
&\text{\Romannum{2}} \quad &&\Delta \bar{p}^{(n+1)}=\nabla \cdot \big(\tilde{\vec{u}}/\Delta t\big), &&\textrm{in} \; \Omega\\
\label{NeumannPressureBC-Euler}
&&&\frac{\partial \bar{p}}{\partial \vec{n}}^{(n+1)}=\vec{n} \cdot\big(\tilde{\vec{u}}/\Delta t\big), &&\textrm{on} \; \partial\Omega\\
\label{UpdatingVelocity-Euler}
&&&\tilde{\tilde{\vec{u}}}=\tilde{\vec{u}}-\Delta t\nabla\bar{p}^{(n+1)}, &&\textrm{in} \; \Omega\\[1em]
\label{UnsteadyStep-Euler}
&\text{\Romannum{3}} \quad && \vec{u}^{(n+1)}=\vec{u}^{(n)}+\tilde{\tilde{\vec{u}}}, &&\textrm{in} \; \Omega.
\end{alignat}
\end{subequations}

\noindent
Here we define the average pressure $\bar{p}^{(n+1)}=1/ \Delta t \int_{t^{n}}^{t^{n+1}} p'/\rho \, dt$
with the normal derivative $\partial \bar{p}^{(n+1)}/ \partial\vec{n}=\vec{n} \cdot \nabla \bar{p}^{(n+1)}$.
To obtain a unique solution for \eqref{Poisson-Euler}-\eqref{NeumannPressureBC-Euler}, we require $\int_\Omega \bar{p}^{(n+1)}=0$.
Note that $p'$ is a correction to the pressure to ensure the divergence-free constraint.
As one notices, the pressure is already present in the stress tensor in equation \eqref{eq:Momentum}.

For the second-order time integration of the velocity, we use the Adams-Bashforth method in the first step \Romannum{1},

\begin{equation*}
\tilde{\vec{u}}_{AB}=\Delta t \sum_{j=0}^{J-1} \beta_j \big(-\vec{u}^{(n-j)}\cdot\nabla\vec{u}^{(n-j)}+\frac{1}{Re} \nabla \cdot (\grvec{\sigma}^{(n-j)}/\rho)+\vec{g}^{(n-j)}/\rho\big),
\end{equation*}

\noindent
with the coefficients $\beta_0=3/2$ and $\beta_1=-1/2$. However, the effective pressure, $\bar{p}^{(n+1)}$, is first-order accurate in time. If required, one can reconstruct the pressure for higher-order accuracy, see \cite{ProjectionAB, Sanderse20123041}. In the above projection scheme, we use by construction $\nabla \cdot \vec{u}^{(n-j)}=0$, for all $j\ge 0$. Therefore, we can integrate the unsteady terms in the third step \Romannum{3}. In this way, by replacing the intermediate velocity from the first step, the right-hand side of the Poisson equation in the second step \Romannum{2} is independent of the time step. This prevents the numerical instability observed in \cite{emamy-tuprints3471, Ferrer2014, Steinmoeller2013480} using the dG method.

\section{Hybrid multiscale method (HMM)}
\label{sec:hmm}

The particle simulations represent a bottle-neck of our hybrid method. To compute the solution of a macroscopic problem, the stress tensor is required at each quadrature point, as shown for example in Figure~\ref{fig:DataTransfer} for a two-dimensional fluid dynamics problem.
In order to reduce the number of these time-consuming simulations, we do not follow the strategy of the one-to-one correspondence between the MD boxes and the quadrature points on the macroscopic level, but we split our simulations into an off-line and on-line phases to prepare approximations for the stress in advance. In an off-line training phase, hence, we collect most of the information using the greedy method (see, e.g., \cite{temlyakov2008greedy}) and build functional dependencies of the stress on shear rates based on the data approximation with Chebyshev's orthogonal polynomials. In an on-line phase of fast multiple queries we obtain  the required data using these approximations and in rare cases rebuild them if new strain rates appear which exceed the approximated intervals.

The stress data is provided by the MD method using one-dimensional shear flow (the so-called simple-shear flow) realized along one of the coordinate axes  as described above.
However, the macroscopic model represents a fluid flow with a two- or three-dimensional velocity field.
In order to obtain one-dimensional shear rates for the particle simulations we have to rotate the MD boxes along the streamlines and change correspondingly the data basis \cite{yasuda2008model}.

Furthermore, the stress data provided by the MD simulations is subject to statistical and systematic errors. To reduce the noise in MD data we apply the method of Proper Orthogonal Decomposition and project our simulation data onto principal component directions.
In what follows we describe our approach in more details.

\begin{figure}[htb]
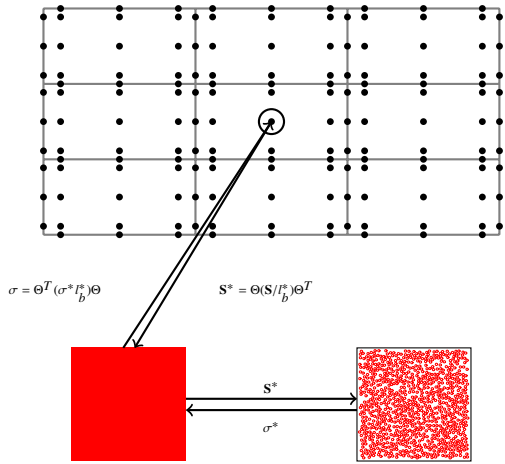

  		\centering
  		\input ./figures/QuadNodes-CFD-MD-withoutS.tex	
  		\caption{Data transfer for the $2$D hybrid simulations. $\Theta$ is the rotational transformation yielding a $1$D strain field.}
  		\label{fig:DataTransfer}
\end{figure}

\subsection{Reduced-order approach for data refinement strategy}
\label{sec:ReducedOrder}
\label{sec:DataRefinement}

In order to find a proper approximation of the stress tensor with fewer number of particle simulations, we solve the optimization problem with a relatively small number of samples. Then we use the greedy algorithm (worst scenario search), see e.g. \cite{temlyakov2008greedy}, for the data refinement to suggest the shear rate(s) for new particle simulations. If one plots the residual $(Ax-b)$ versus $\dot{\gamma}$, the proposed shear rate for a new simulation  $\dot{\gamma}_{new}=0.5(\dot{\gamma}_{M}+\dot{\gamma}_{N})$, is found in the neighborhood of $\dot{\gamma}_{M}=\argmax_{\dot{\gamma}}(Ax-b)$, where $\dot{\gamma}_{N}$ is the left or right neighbor of $\dot{\gamma}_{M}$ which corresponds to the larger residual.

\subsection{Eigenvalue decomposition of strain and stress fields}

Any strain rate tensor can be written as a symmetric matrix of streaming velocity gradients
$
\mS=(1/2)(\nabla \matr{u}+ \nabla \matr{u}^T)
$.
In the case of two-dimensional simulations the strain rate matrix is

\begin{equation}
\mS=
\begin{pmatrix}
\frac{\partial u}{\partial x} & S_{12} \\
S_{12} & \frac{\partial v}{\partial y}
\end{pmatrix},
\label{eq:mS2d}
\end{equation}

\noindent
where $S_{12} = \left(\frac{\partial u}{\partial y}+\frac{\partial v}{\partial x}\right)/2$.
In the case of plane flow in three-dimensional simulations
the strain rate matrix is

\begin{equation}
\mS=
\begin{pmatrix}
\frac{\partial u}{\partial x} & 0 & S_{13} \\
0 & 0 & 0 \\
S_{13} & 0 & \frac{\partial w}{\partial z}
\end{pmatrix}.
\label{eq:mS3d}
\end{equation}

\noindent
where $S_{13}=\left(\frac{\partial u}{\partial z} + \frac{\partial w}{\partial x}\right)/2$. 

In the following part, for simplicity we focus on the derivation of the basis transform for the two-dimensional case. The plane flow in the three-dimensional case follows analogously and in the next section we also compare flow profiles of 2D Couette and 3D Poiseuille flows with analytic solutions as proof of concept.

As pointed out in \cite{Todd1998} there exists an angle $\theta=\frac{1}{2}\arctan(-\frac{S_{11}}{S_{12}})$,
where $S_{ij}$ are components of the strain rate matrix $\mS$ in (\ref{eq:mS2d}), and a rotation matrix

\begin{equation}
\mTheta=
\begin{pmatrix}
\cos\theta & \sin\theta \\
-\sin\theta & \cos\theta
\end{pmatrix},
\end{equation}

\noindent
which transforms the strain rate tensor $\mS$ to the anti-diagonal matrix

\begin{equation}
\mS'=\mTheta\mS\mTheta^T=
\begin{pmatrix}
0 & \dgamma/2 \\
\dgamma/2 & 0
\end{pmatrix}\ .
\label{eq:mSprim}
\end{equation}

This strain rate tensor $\mS'$ corresponds to a pure-shear deformation (i.e., in absence of normal stresses) with the shear rate
$\dgamma/2=-S_{11}\sin(2\theta) + S_{12}\cos(2\theta)$.
Eigenvalue decomposition of the pure-shear strain rate matrix $\mS'$ yields

\begin{equation}
\mS'=\mP\mLam\mP^T
=
\mP
\begin{pmatrix}
\lambda_1 & 0 \\
0 & \lambda_2
\end{pmatrix}
\mP^T\ ,
\end{equation}

\noindent
where $\lambda_i\in\raumR$  are the eigenvalues of $\mS'$, $\mP=(\matr{p}_1,\matr{p}_2)$
is the matrix of the corresponding eigenvectors.
By solving the eigenvalue equation $\mS'\mP=\lambda\mP$ one can find that $\lambda_{1,2}=\mp{\dgamma}/2$ and
from $\mS'\matr{p}_i=\lambda\matr{p}_i$ it follows that
$\mP=\frac{1}{\sqrt{2}}
\begin{pmatrix}
-1 & 1 \\
1 & 1
\end{pmatrix}.
$

Thus, the eigenvalue decomposition of $\mS'$ reads

\begin{equation}
\mS' =
\frac{1}{2}
\begin{pmatrix}
-1 & 1 \\
1 & 1
\end{pmatrix}
\begin{pmatrix}
-\dgamma/2 & 0 \\
0 & \dgamma/2
\end{pmatrix}
\begin{pmatrix}
-1 & 1 \\
1 & 1
\end{pmatrix}
\left[
=
\begin{pmatrix}
0 & \dgamma/2 \\
\dgamma/2 & 0
\end{pmatrix}
\right]
\end{equation}

\noindent
and is same as (\ref{eq:mSprim}).

We are now looking for a fitting function $f(\dgamma)$ approximating the MD stress data $\msigma^{\rm MD}\equiv\{\sigma_{\alpha\beta}(\dgamma)\}_{N_{\dgamma}}^{N_{\rm sets}}$ for $N_{\dgamma}$ values of the shear rate ${\dgamma}$ and ${N_{\rm sets}}$ independent data sets

\begin{equation}
f(\dgamma)
\approx \msigma^{\rm MD}\ .
\end{equation}

\noindent
Thus, $f(\mS')$ should approximate $\msigma^{\rm MD}$

\begin{equation}
\label{eq:fe}
f(\mS')
=
f\left(\mP
\begin{pmatrix}
\lambda_1 & 0 \\
0 & \lambda_2
\end{pmatrix}
\mP^T
\right)
=
\mP
\begin{pmatrix}
f(\lambda_1) & 0 \\
0 & f(\lambda_2)
\end{pmatrix}
\mP^T
\approx \msigma^{\rm MD}.
\end{equation}


\noindent
Substituting the eigenvalues of $\lambda_{1,2}$ into (\ref{eq:fe}) yields

\begin{equation}
\msigma^{\rm MD}
\approx
\mP
\begin{pmatrix}
	f(-\dgamma/2) & 0 \\
	0 & f(\dgamma/2)
\end{pmatrix}
\mP^T.
\label{eq:sigmamd}
\end{equation}

This means that in order to obtain a least-square approximation $f(\dgamma)$ we need to transform MD data into the eigenvector basis $\mP$

\begin{equation}
\begin{pmatrix}
f(-\dgamma/2) & 0 \\
0 & f(\dgamma/2)
\end{pmatrix}
\approx
\mP^{T}\msigma^{\rm MD}\mP=
\begin{pmatrix}
\tilde{\sigma}_{11} & \tilde{\sigma}_{12}  \\
\tilde{\sigma}_{21} & \tilde{\sigma}_{22}
\end{pmatrix}
\equiv \tsigma.
\label{eq:tsigma}
\end{equation}

We can therefore assume $\tilde{\sigma}_{12}=\tilde{\sigma}_{21}=0$ and fit

\begin{equation}
f(\lambda_1)\equiv f(-\dgamma/2)\approx\tilde{\sigma}_{11}
\end{equation}
\label{eq:flam1}
\begin{equation}
f(\lambda_2)\equiv f(\dgamma/2)\approx\tilde{\sigma}_{22}
\label{eq:flam2}
\end{equation}

The stress data, $\msigma^{\rm MD}$, were obtained  for various densities  
using the SLLOD method with a WCA potential in two- and three-dimensional MD simulations, as discussed in Section \ref{sec:md}.
In Figure~\ref{fig:mdsigma} we show 2D MD stress data at $\rho=0.8$ and in Figure~\ref{fig:princ} the results of the orthogonal transform of this data into the eigenvector basis representation, $\tsigma$,  via (\ref{eq:tsigma}). The diagonal principal component functions $\tsigma_{11}(\lambda_1)$ and $\tsigma_{22}(\lambda_2)$ look very simple: $\tsigma_{11}$ for $\lambda_1\le0$ is nearly a straight line, $\tsigma_{22}$ for $\lambda_2\ge0$ increases monotonously with a negative curvature, which changes with the density. 
These features make it easier to approximate $\tsigma_{11}$ and $\tsigma_{22}$ by low order polynomials $f(\lambda_1)$ and $f(\lambda_2)$, respectively. We also consider an approximation of the principal stress by $f(\lambda)$ with $\lambda=\lambda_1\cup\lambda_2$.

Note that the off-diagonal components of the matrix at the left-hand-side of (\ref{eq:tsigma}) are zero. However, due to statistical errors of MD simulations (and perhaps due to a systematic error of the method) the off-diagonal components $\tilde{\sigma}_{12}$ and $\tilde{\sigma}_{21}$ of the matrix $\tsigma$ deviate from zero. These stochastic and systematic deviations increase with the shear rate, decrease with the density and lay in the range of few percents if compared to the values of the diagonal principal stress components. As mentioned above, we suppress these deviations by setting $\tilde{\sigma}_{12}=\tilde{\sigma}_{21}=0$ in the following processing of our data.

\begin{figure}[htbp]
\subfigure{
\includegraphics[width=0.5\textwidth, trim=0.5cm 0.5cm 0cm 1cm, clip]{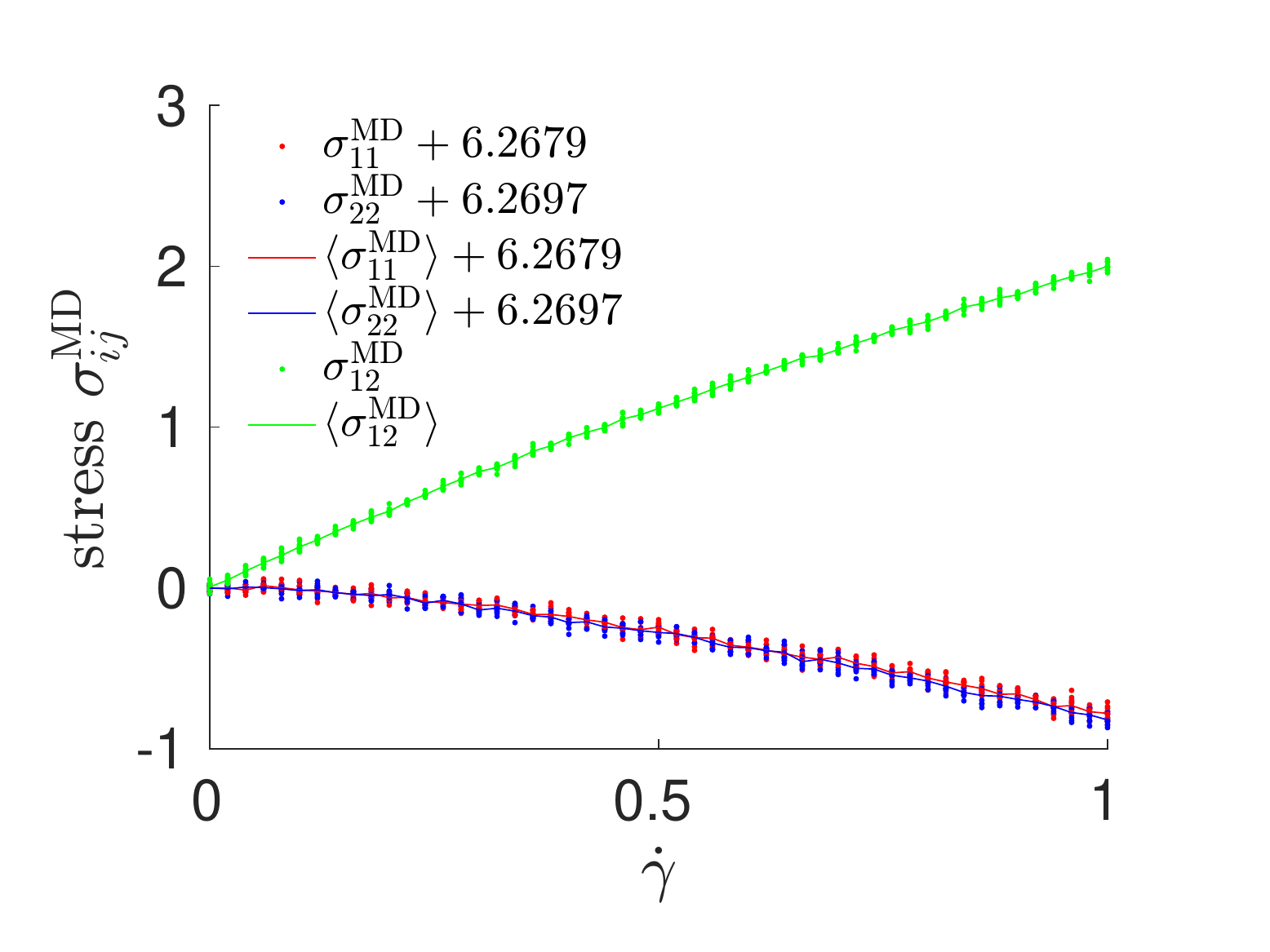}
}
\caption{
	MD results of a stress tensor, $\msigma^{\rm MD}$, in a simple-shear 2D flow using the SLLOD method with a WCA potential, at the density $\rho=0.8$. Points are simulation data, lines are mean averages of the simulation data. The normal stress components, $\sigma^{\rm MD}_{ii}(\dot\gamma)$, are shifted by a constant value of the zero-shear mean stress, $\langle\sigma^{\rm MD}_{ii}({\dot\gamma}=0)\rangle=-6.2679$.}
\label{fig:mdsigma}
\end{figure}

\begin{figure}[htbp]
\subfigure{
\includegraphics[width=0.5\textwidth, trim=0cm 0cm 0cm .6cm, clip]{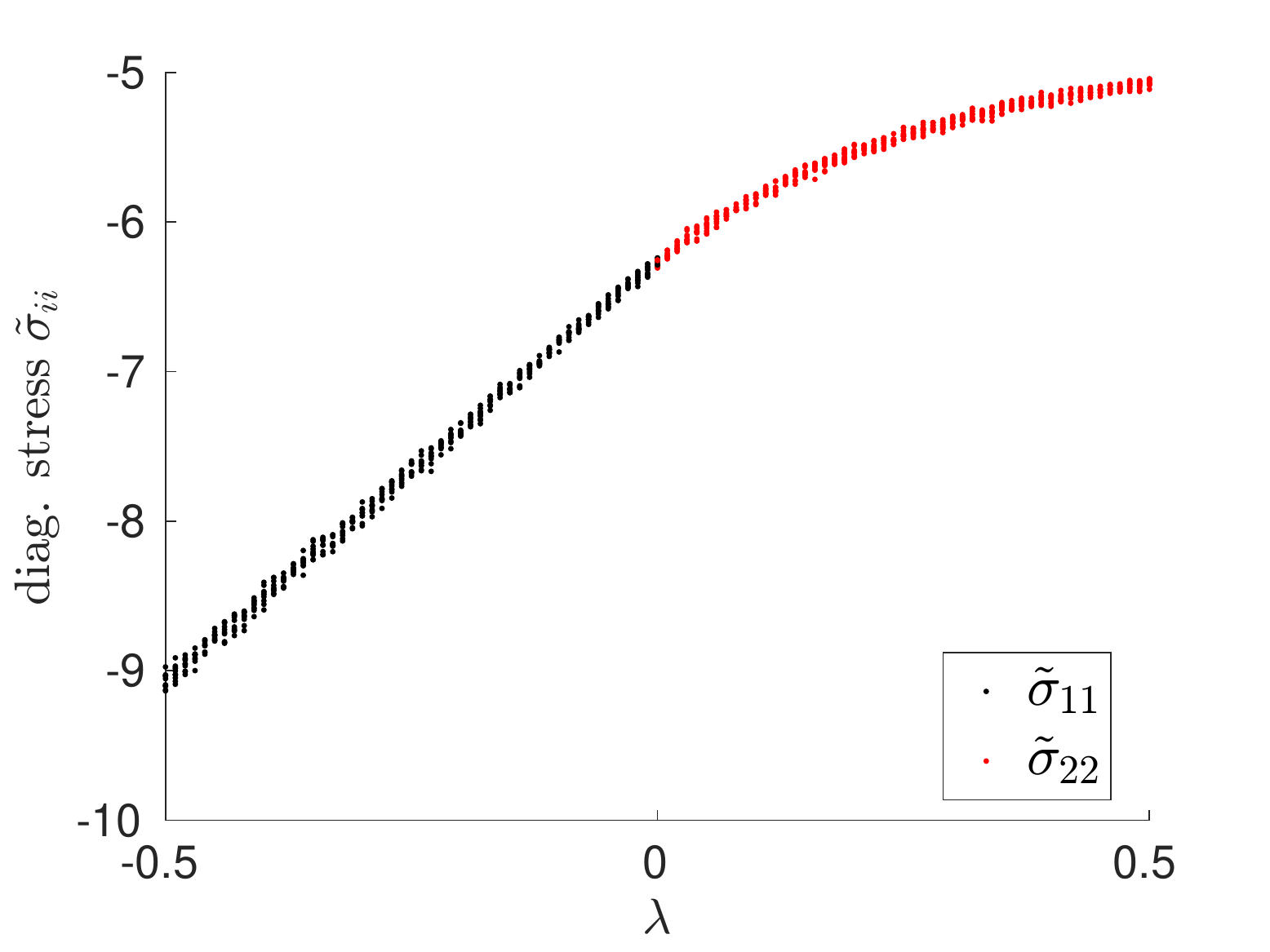}
}
\subfigure{
\includegraphics[width=0.5\textwidth, trim=0cm 0cm 0cm .6cm, clip]{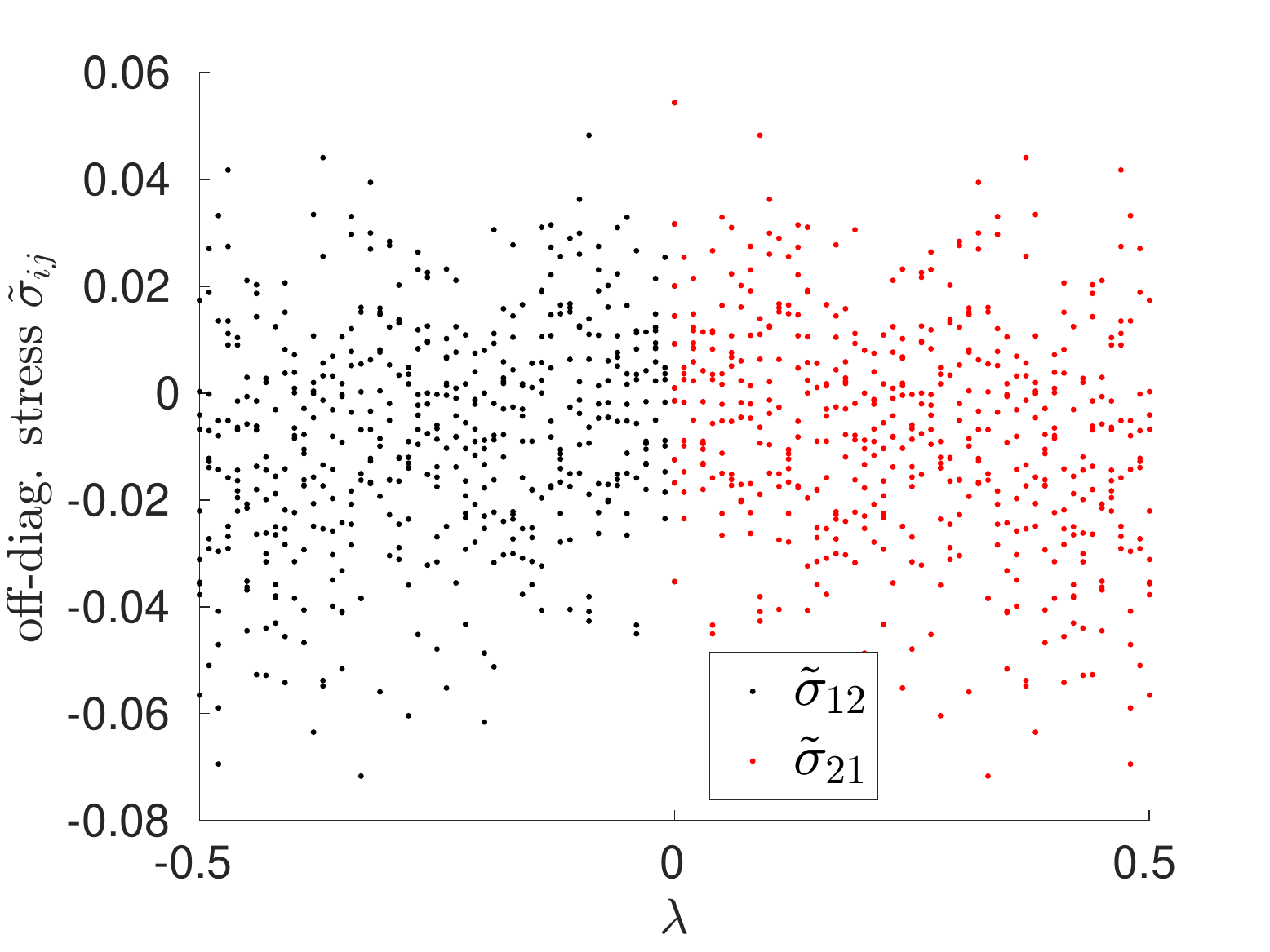}
}
\captionof{figure}{
	Diagonal (top) and off-diagonal (bottom) principal stresses, $\tsigma$, obtained by the orthogonal transform \eqref{eq:tsigma} of the molecular stress data, $\msigma^{\rm MD}$,  shown in Figure~\ref{fig:mdsigma}.
}
\label{fig:princ}
\end{figure}

\subsection{Proper Orthogonal Decomposition for noise reduction}

In order to extract physically significant information from our MD data and to reduce the noise we apply the Proper Orthogonal Decomposition (POD) method,  which is based on a Singular Value Decomposition (SVD) of a matrix. Hence, the matrix of principal stress can be represented as
$\tsigma=\mV\mSigma\mZ^T\label{eq:svd}$,
where $\mSigma$ is a diagonal $m\times n$ matrix of rank $r\le \min(m,n)$ with  positive singular values $s_1\ge s_2\ge s_r>0$ stored  along the main diagonal

\begin{equation}
\mSigma=
\begin{pmatrix}
s_1&&0\\
&\ddots\\
0&&s_r\\
&&\\
&0&\\
&&
\end{pmatrix},
\label{eq:mSigma}
\end{equation}

\noindent
$\mV$ is a $m\times m$ unitary matrix of left-singular vectors, $\mZ$ is a $n\times n$ unitary matrix of right-singular vectors. 
$m$ is the number of $\dgamma_i$'s, $n$ is the number of independent simulations sampling $\msigma^{\rm MD}(\dgamma_i)$. 
The singular values $s_i$ are related to the eigenvalues of the correlation matrix $\mC=\tsigma^T\tsigma$ as $s_i^2=\lambda_i^{(\mC)}$.
According to the POD method one can reduce the rank of the original stress matrix $\tsigma$ by a low-rank approximation $\tsigma_{(k)}$ by keeping the first $k$ of $r$ singular values $s_i$ in $\mSigma$

\begin{equation}
\tsigma \approx \tsigma_{(k)} = \mV\mSigma_{(k)}\mZ^T = \sum\limits_{i=1}^k s_i\mV_i \otimes\mZ_i^T\ ,
\label{eq:tsigmak}
\end{equation}

\noindent
where $k$ is a reduced rank, $\mSigma_{(k)}$ is a reduced-rank singular-value matrix.

The SVD analysis of our principal stresses has shown that there exists one singular value, $s_1$, which is by two orders of magnitude greater than any other $s_i$, as shown in Figure~\ref{fig:pod_fl_det-rho0.8} for 10 independent 2D MD data sets. The remaining singular values $s_i$ with $(i>1)$ have nearly same amplitude and correspond to the statistical noise of the MD data. Therefore, the first principal component with the largest variance along the principal direction given by the vector $\mV_1$ approximates our principal stress data

\begin{equation}
\tsigma \approx \tsigma_{(1)} = s_1\mV_1 \otimes\mZ_1^T
\label{eq:tsigma_1}
\end{equation}

\noindent
with the projection relative error

\begin{equation}
\epsilon^2_{(1)}= \frac{ \sum\limits_{i=2}^{r} s_i^2} {\sum\limits_{i=1}^{r} s_i^2}.
\end{equation}

\noindent
For instance, in the case of singular values shown in Figure~\ref{fig:pod_fl_det-rho0.8}, the projection relative error of the rank-1 approximation,  $\tsigma_{(1)}$, is $\epsilon_{(1)} \lessapprox 0.9 \%$.

\begin{figure}[htbp]
\subfigure{
\includegraphics[width=0.5\textwidth, trim=0cm 0cm 0cm 1.75cm, clip]{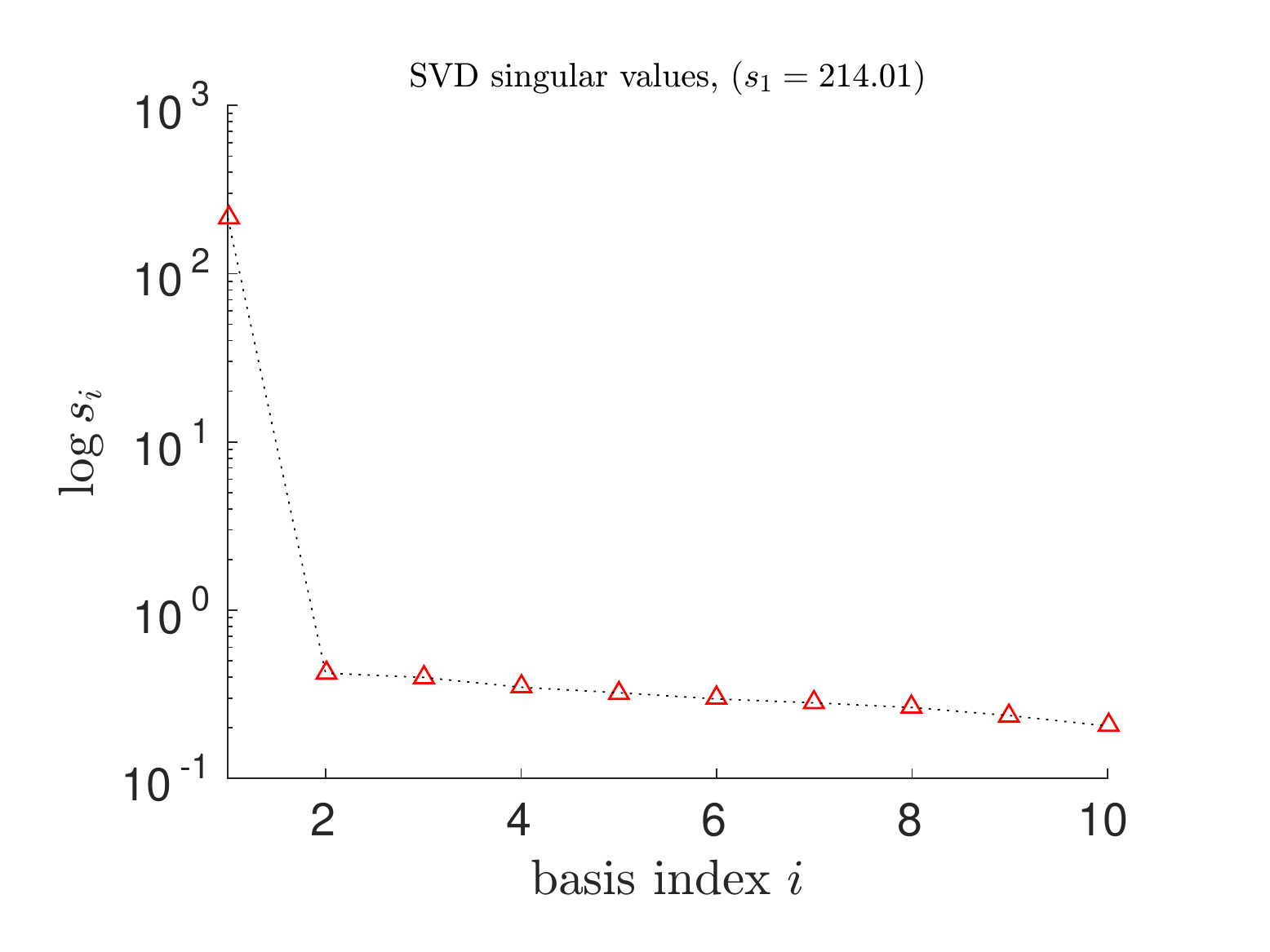}
}
\captionof{figure}{
	Singular values of the principal stress at the density $\rho=0.8$ obtained using the SVD method.}
\label{fig:pod_fl_det-rho0.8}
\end{figure}

\subsection{Chebyshev's approximation with Tikhonov's regularisation}

The noise-reduced data of the principal stress, $\tsigma_{(1)}$, will be now  approximated using the least squares method with the orthogonal polynomials of Chebyshev

\begin{equation}
\tsigma_{(1)}\approx f(\lambda) =
\sum\limits_{i=0}^{k} a_i T_i(\lambda)\ ,\quad \lambda \in \raumR\ .
\end{equation}

\noindent
Here $k$ is the degree of the approximating polynomial, $a_i$ coefficients of the Chebyshev polynomials $T_i$.
Our goal is to approximate simulation data by means of a simple function for $f(\dgamma)$ for further use at the level of the macroscopic solver.
To reduce a large number of particle simulations we use an off-line training phase and an on-line phase of fast multiple queries. For this training, we solve a least-square problem with the Tikhonov regularization for each principal component of the stress tensor. Tikhonov's regularization improves the approximation of a badly conditioned data matrix, i.e., when the problem is ill-posed.
Thus, our goal is to find a vector $\mx$ minimizing an extended residuum functional

\begin{equation}
\label{eq:LeastSquare}
\argmin_{\mx} \big(\llrrvline{\mA\mx-\mb}_2^2+\alpha_1^2\llrrvline{\mx}_2^2+\alpha_2^2\llrrvline{\mD\mx}_2^2\big).
\end{equation}

\noindent
Here $\mb$ is the vector of $n$ ($n \ge k+1$) data points of the corresponding component of the principal stress tensor, $\tsigma_{(1)}$, obtained as described in the previous section.
Further, $\mx$ is the vector of the unknown Chebyshev coefficients $a_i$.
$\mA$ is the Vandermonde matrix of the Chebyshev polynomial basis.
The penalty term $\mD\mx$ is added to damp the oscillations in the derivative function $\partial \mA /\partial \dgamma \,\mx=\mA\mD\mx$.
Parameters $\alpha_1$ and $\alpha_2$ are the regularization parameters.
The equivalent problem $(\mA^T\mA+\alpha_1^2\mI+\alpha_2^2\mD^T\mD)\mx=\mA^T\mb$ is solved by the LU factorization using the LAPACK library \footnote{http://www.netlib.org/lapack}.

The Chebyshev approximation with the Tikhonov regularization is shown in Figure~\ref{fig:tikh-sum}.
As it has been mentioned, the dependence  $f(\lambda)$ exhibits two distinct behaviour: it is nearly linear for $\lambda<0$ and is concave for $\lambda>0$ with a density dependent curvature. Therefore, we approximate $f(\lambda)$ either by one joint function in the range $\lambda \in \raumR$ or by two separate functions $f(\lambda_1)$ and $f(\lambda_2)$ with $\lambda_i$ in ranges $\lambda_1<0$ and $\lambda_2>0$. In both cases, the joint and split approximations are indistinguishable in the plot scales, however, the degree of the optimal polynomials for the joint approximation $f(\lambda)$ is higher than that for the split approximations $f(\lambda_i)$. From the practical point of view a low degree polynomial approximation is preferred, since it can significantly improve the numerical efficiency and stability of the automated algorithms for data processing.

\begin{figure}[htbp]	
\subfigure{
\includegraphics[width=0.5\textwidth, trim=0.5cm 0.5cm 0cm 0cm, keepaspectratio, clip]{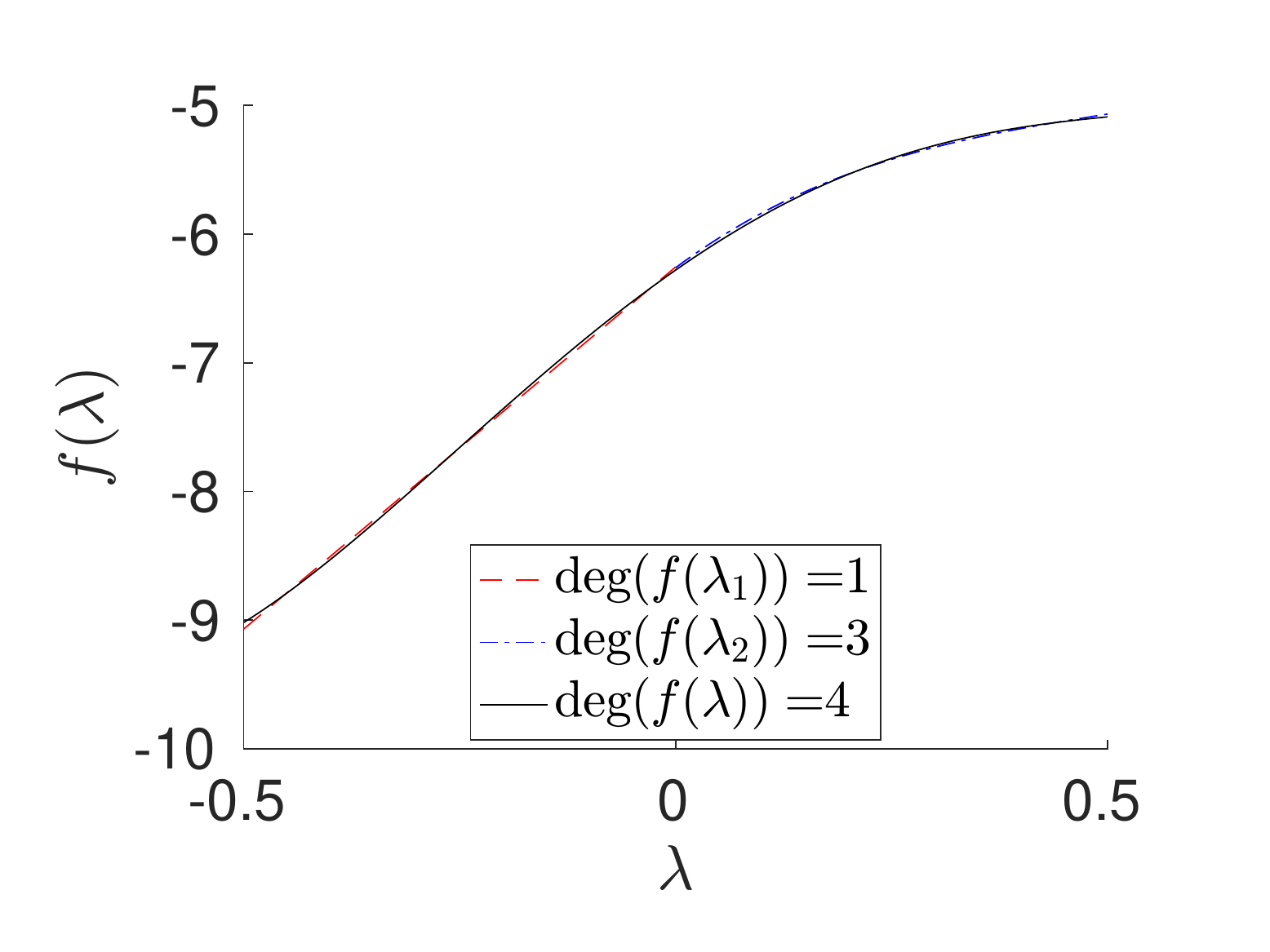}
}
\captionof{figure}{
	Approximations of the principal stress by the Chebyshev polynomials using either one joint function $f(\lambda)$ in the range $\lambda \in [-0.5, 0.5]$ or two functions $f(\lambda_1)$ and $f(\lambda_2)$ with $\lambda_i$ in ranges $\lambda_1 \in [-0.5,0]$ and $\lambda_2 \in [0, 0.5]$.
	The approximation degree is shown in the legends.
}
\label{fig:tikh-sum}
\end{figure}	

\subsection{Back transform of Chebyshev's approximants}

On the macroscopic level, the stress data is requested from the MD simulations level at each quadrature point for a given strain rate, $\mS$. 
Hence, the principal stress calculated by the Chebyshev approximations, $f(\lambda_i)$ with $\lambda_1=-\dgamma/2$, $\lambda_2=\dgamma/2$, has to be transformed back to the basis space of the macroscopic solver, $\msigma(\mS)$.
Using (\ref{eq:mSprim}), (\ref{eq:sigmamd}) and the property of the Galilean invariance of the rotation $\mTheta$ we have

\begin{equation}
\msigma(\mS)=\mTheta^T\mP
\begin{pmatrix}
f(-\dgamma/2) & 0\\
0& f(\dgamma/2)
\end{pmatrix}
\mP^T\mTheta\ .
\label{eq:msigmaback}
\end{equation}

The rotation matrix $\mTheta$ is problem specific. 
Hence, in Figures~\ref{fig:back} we compare the MD stress data, $\sigma_{ij}^{\rm MD}(\dgamma)$, with $\msigma^{\rm CHEB}(\dgamma)$, the back transform of $f(\lambda_i)$ to the MD basis space obtained from (\ref{eq:sigmamd}).
One can see that the shear stress, $\sigma_{12}^{\rm MD}(\dgamma)$, is described very well.
Some discrepancies can be observed for the normal stresses: $\sigma_{11}^{\rm MD}$ and $\sigma_{22}^{\rm MD}$ from MD exhibit slight normal stress difference at $\rho=0.8$, whereas $\msigma^{\rm CHEB}_{11}$ and  $\msigma^{\rm CHEB}_{22}$ are identical (in Figure~\ref{fig:back}, only $\msigma^{\rm CHEB}_{11}$ is shown). 
This is a consequence of the suppression  in (\ref{eq:tsigma}) of the off-diagonal components, $\tilde{\sigma}_{12}$ and $\tilde{\sigma}_{21}$. For $\rho=0.8$ these terms are nearly zero and, hence, these discrepancies can hardly be recognized. 
Furthermore, we have checked the descriptive quality of the back transformation of the joint  approximation, $f(\lambda)$ with $\lambda = \lambda_1 \cup \lambda_2$, which includes higher degree Chebyshev's polynomials.
It yields very similar results, as one could expect after comparison in Figure~\ref{fig:tikh-sum}.

\begin{figure}[htbp]	
\subfigure{
\includegraphics[width=0.5\textwidth, trim=0.5cm 0.5cm 0cm 0cm, keepaspectratio, clip]{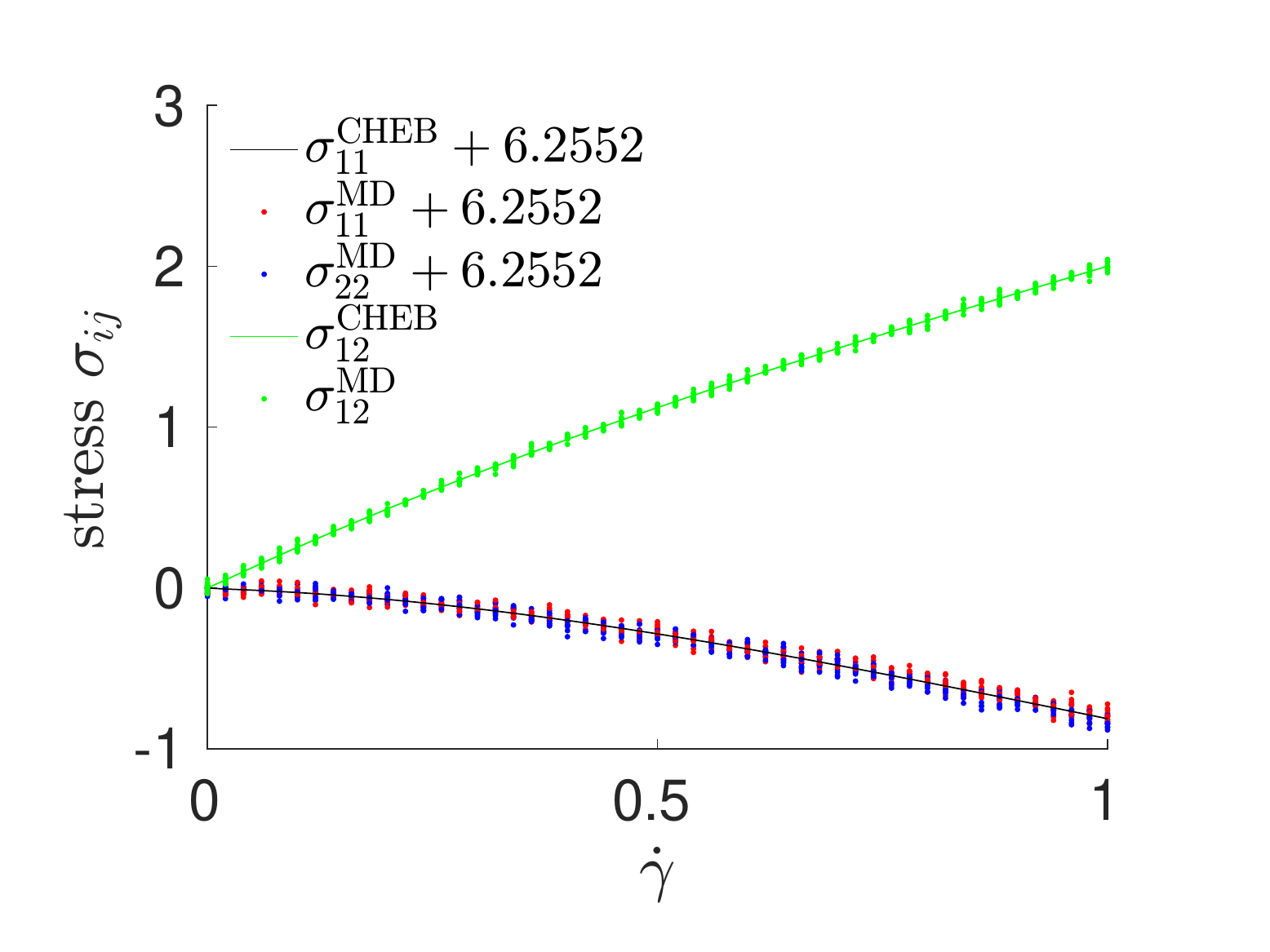}}
\captionof{figure}{
	Components of the 2D stress tensor, $\msigma(\dgamma)$, obtained by the back transform to the MD basis using (\ref{eq:sigmamd}) of the Chebyshev approximation to the principal stress, $f(\lambda_i)$.
	The two-function representation of $f(\lambda_i)$ is used with $\lambda_1= -\dgamma/2$ and $\lambda_2 = \dgamma/2$. The normal stress components, $\sigma_{ii}(\dot\gamma)$, are shifted by a constant value of the zero-shear mean stress, $\langle\sigma^{\rm CHEB}_{ii}({\dot\gamma}=0)\rangle=-6.2552$.
	Lines: Chebyshev's approximation, symbols: MD stress data.
}
\label{fig:back}
\end{figure}

\section{Reduced-order hybrid simulations}


The macroscopic hybrid simulations of the Couette flow are performed on the domain $[-1,1]\times[0,1]$.
The flow is periodic in the streamwise $x$-direction. The no-slip boundary condition is applied at the walls.
At the lower wall $y=0$, the velocity is zero. At the upper wall $y=1$, the velocity in $x$-direction is equal
to $U$ and the velocity in $y$-direction is zero.
A grid of $3 \times 3$ cells is employed. A polynomial degree $k=1$ is assigned in the dG method, $Re=1$.
The velocity profiles presented in Figure~\ref{fig:Couette-2D} overlap with the analytical solution $u_x=yU$
for the equivalent Newtonian fluid with density $\rho=0.8$ and viscosity $\mu=\left.\sigma_{12}/\dgamma\right|_{\dgamma=1}=1.96481$ calculated from the 2D MD data.


\begin{figure}
\includegraphics[width=0.45\textwidth, keepaspectratio, clip]{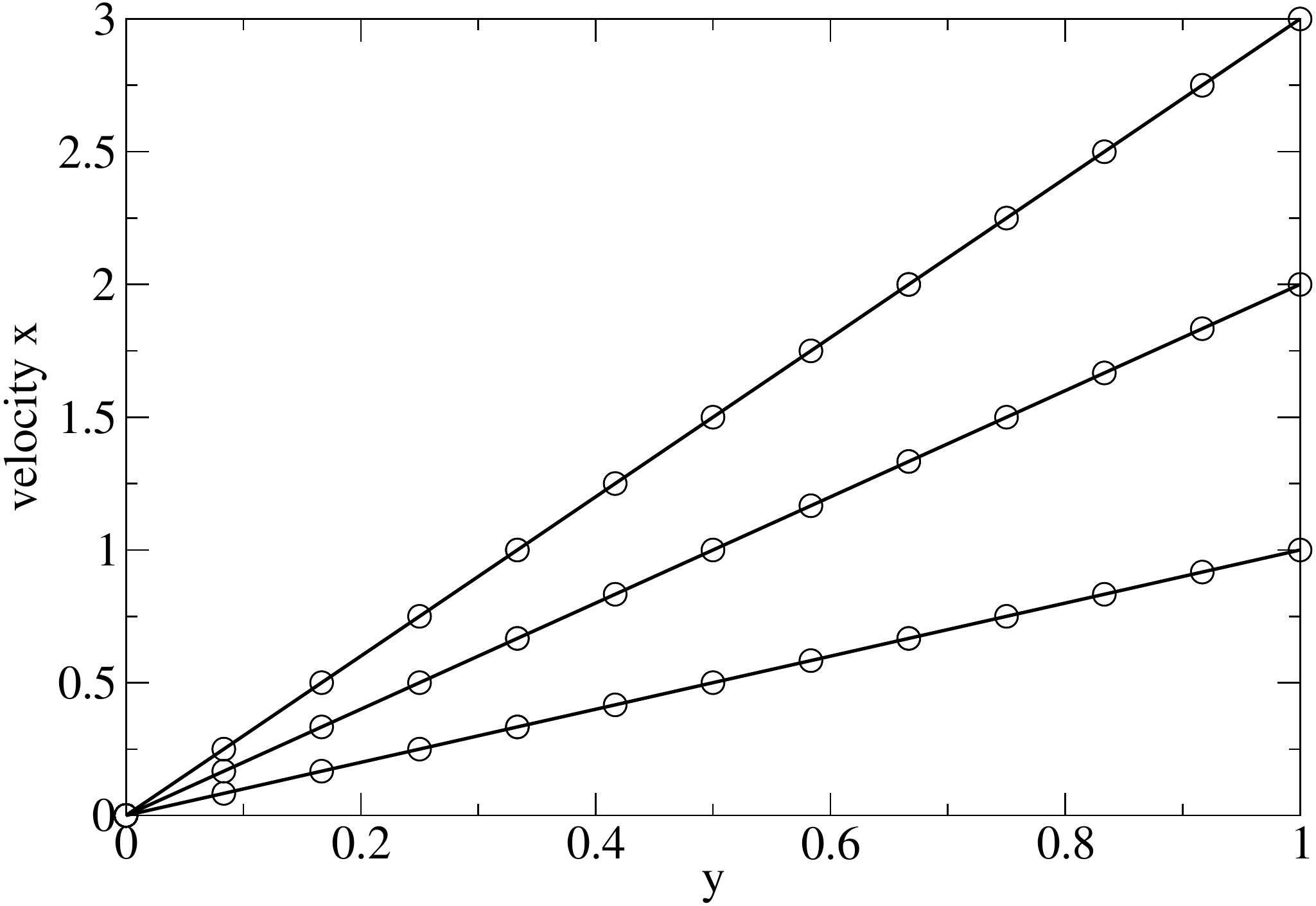}
\caption{2D Couette flow: velocity profiles in the streamwise direction for shear velocities $U=1,2,3$. Symbols: hybrid simulations, lines: analytic solutions $u_x=yU$.}
\label{fig:Couette-2D}	
\end{figure}

\begin{figure}[htbp]	
	\subfigure{
		\includegraphics[width=0.5\textwidth, trim=0.5cm 0cm 0cm 1.6cm, keepaspectratio, clip]{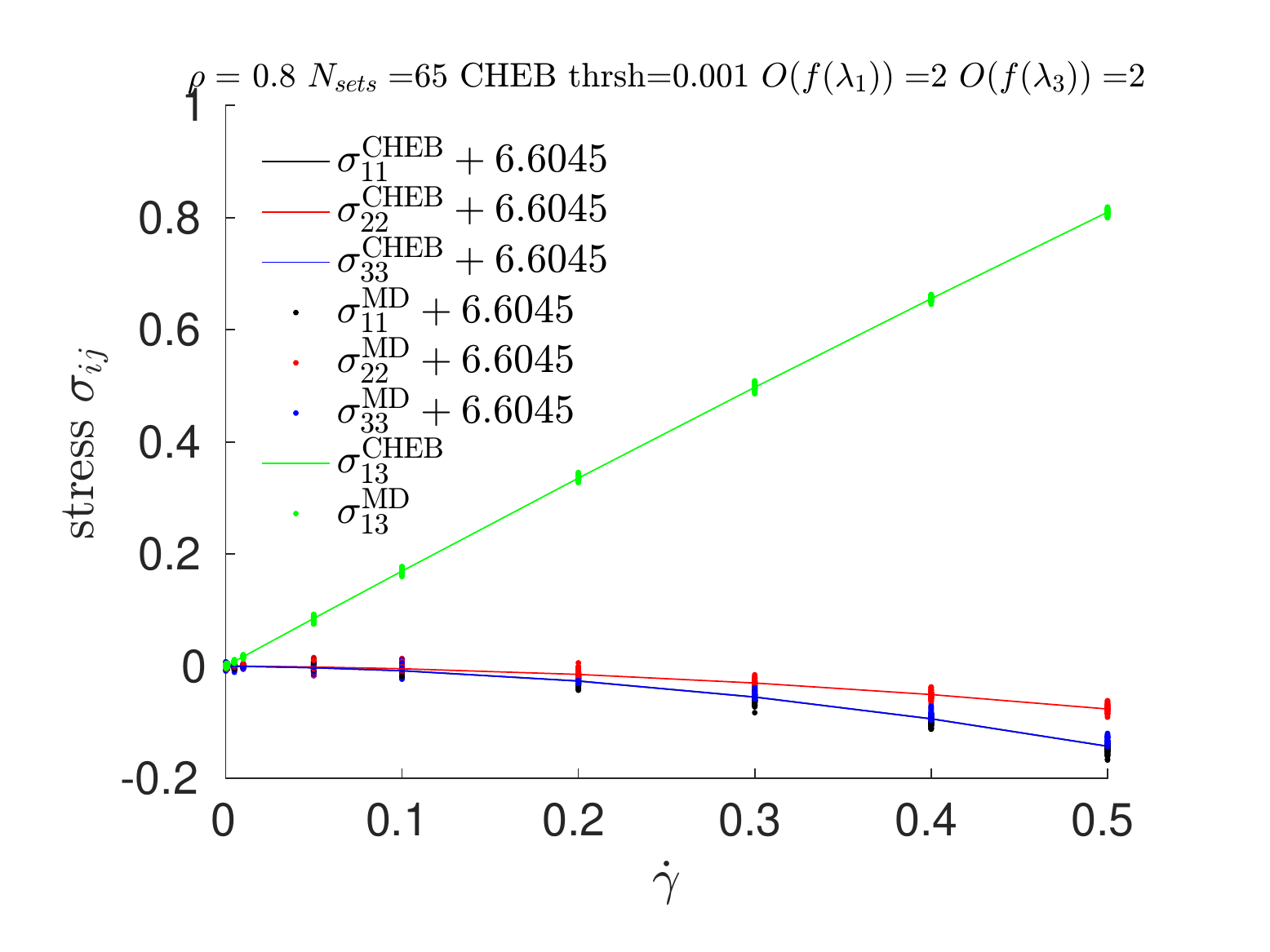}}
	\captionof{figure}{
		Components of the 3D stress tensor, $\msigma(\dgamma)$, obtained from the Chebyshev approximation to the principal stress and its back transform to the MD basis.
		The normal stress components, $\sigma_{ii}(\dot\gamma)$, are shifted by a constant value of the zero-shear mean stress, $\langle\sigma^{\rm CHEB}_{ii}({\dot\gamma}=0)\rangle=-6.6045$.
		Lines: Chebyshev's approximation, symbols: MD stress data.
	}
	
	\label{fig:back3D}
\end{figure}

\begin{figure}[h!] 
\includegraphics[width=0.45\textwidth, keepaspectratio, clip]{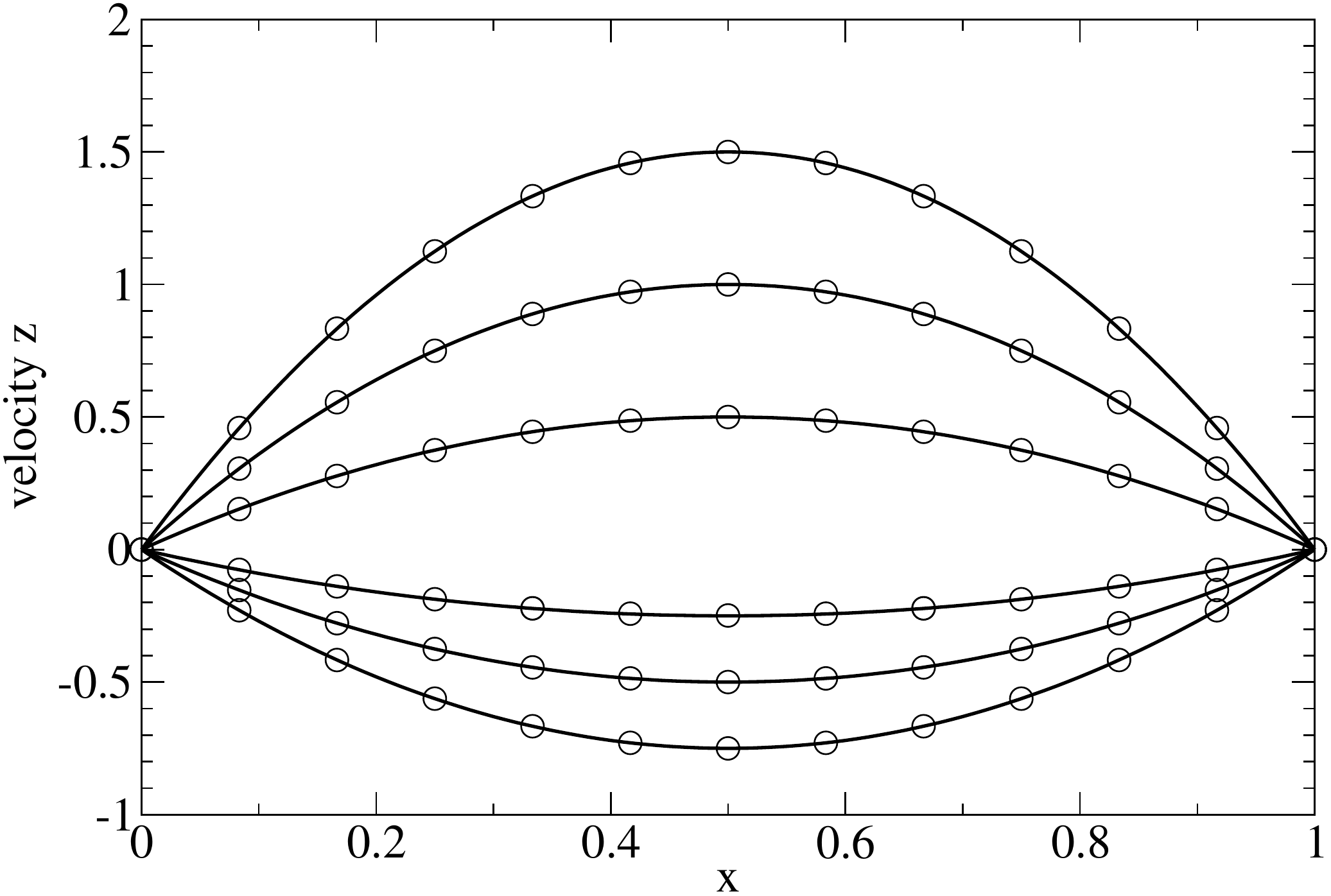}
\caption{3D Poiseuille flow: velocity profiles in the streamwise direction for pressure gradient parameter $P=-3,-2,-1,2,4,6$.  Symbols: hybrid simulations, lines: analytic solutions $u_z=x(1-x)P$.}	
\label{fig:Poisseuille-3D}
\end{figure}

\begin{figure}
	\includegraphics[width=0.5\textwidth, trim=2.2cm 2.6cm 2.2cm 3.6cm, keepaspectratio, clip]{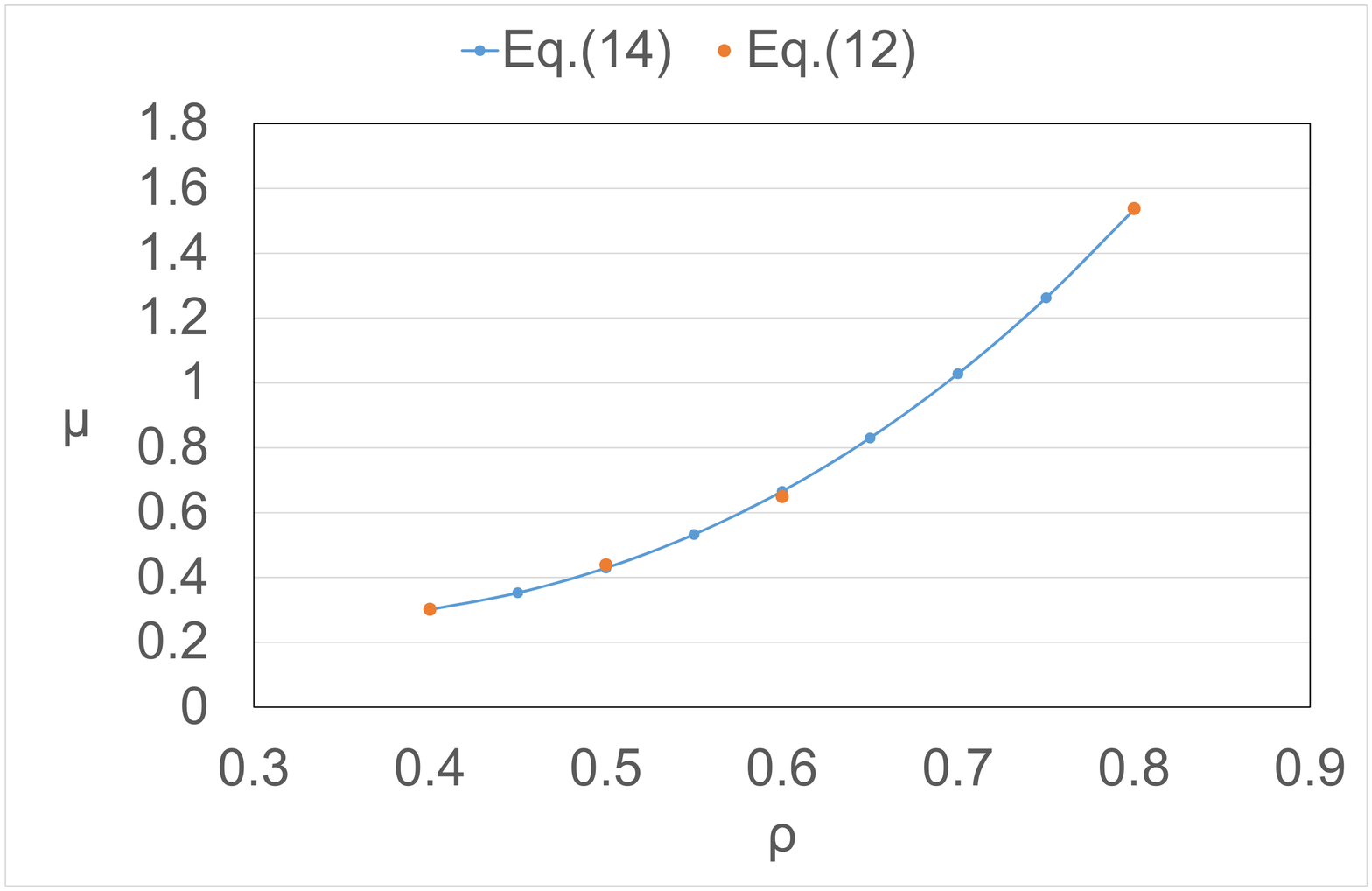}
	\captionof{figure}{Estimated Newtonian fluid viscosity vs. density in $3$D MD.}
	\label{fig:Viscosity-3D}
\end{figure}

Three-dimensional hybrid simulations of the Poiseuille flow are performed for $\rho=0.8$ using the Chebyshev approximations to the $3$D MD data, analogously to the 2D case described in the previous section.
Figure~\ref{fig:back3D} shows the MD data and the approximations for the 3D case. 
In Figure~\ref{fig:Poisseuille-3D} velocity profiles in $z$-direction are plotted for different values of the pressure gradient parameter $P$ and coincide with the analytic solution $u_z=x(1-x)P$.
We take the shear stress $\sigma_{13}$ to estimate the viscosity of the equivalent Newtonian fluid using relation
$\mu=\left.\sigma_{13}/\dgamma\right|_{\dgamma=0.5}=1.61776$.
The viscosities for different densities are plotted in Figure~\ref{fig:Viscosity-3D}, cf.~\cite{emamy_2017}.
The computational domain is taken as $[0,1]\times[-1,1]\times[-1,1]$. The flow is periodic in $y$- and $z$-directions.
The no-slip boundary condition is applied at the walls $x=0$ and $x=1$, where the velocity is zero.
The pressure gradient $f_z=-2\mu\,P/Re$ is applied in the streamwise $z$-direction with $Re=1$.
A grid of $3\times 3\times 3$ cells is employed. The polynomial degree is $k=2$ in the dG method.

\section{Conclusion}

In hybrid particle continuum schemes, molecular simulations typically pose a computational bottleneck.

In the first part of this paper we discuss in detail strategies to minimize the effort of particle-based simulations. We find that it is preferable to simulate small system sizes and determine limits of the latter, which are imposed by finite-size effects such as the transition to two-dimensional behavior.

As we are interested in properties of the steady-state and not necessarily dynamics, a simple non-physical thermostat such as the isokinetic thermostat with SLLOD boundary conditions is actually preferable while thermostats such as Lowe-Andersen allow for simulations at elevated viscosity. We also investigate boundaries for the size of time steps, and discuss the mapping of time scales between micro- and the macroscale.

In the second part we propose a novel reduced-order technique, which allows us to approximate the stress-strain relation in efficient way and thereby further reduce the computational effort of the microscopic simulations. Our approach is based on a delicate combination of the eigenvalue decomposition, the least square approximation using the Chebyshev polynomials with the Tikhonov regularization and the proper orthogonal decomposition. The latter is applied to reduce statistical noise from the MD simulation data.
 Our hybrid dG-MD method is tested by evaluating velocity profiles of Couette and Poiseuille flow. To our knowledge this is the first time, such  reduced-order techniques are applied in the context of hybrid heterogeneous atomistic-continuum simulations.

\section*{Acknowledgments}

The present work is supported by German Science Foundation (DFG) under the grant TRR 146 (project C~5).
We would like to thank M.~Oberlack, F.~Kummer and B.~M\"{u}ller for providing their code BoSSS and T.~Raasch for fruitful discussions on the topic.
The authors gratefully acknowledge the computing time granted on the HPC cluster Mogon at Johannes Gutenberg-University Mainz.

\appendix
\section{Mapping}
\label{app:sec:Mapp}
We assume that our particles correspond to colloidal particles. The typical size of a colloid is 1 $\mu m$ (range 100 $nm$ to 10 $\mu m$). Thus, the unit of length in the MD simulations is $\sigma = 10^{-6} ~m$.

The time scale in the simulation is given by the so-called MD time $t_{MD}$, corresponding to $\sim 10^4$ time steps in the simulation (for a time step of $10^{-4}$).
\begin{align*}
t _{MD} &=\sigma \cdot \sqrt{\frac{m}{\varepsilon}}\\
\varepsilon &= 1 k_BT\\
&= 1.3\cdot 10^{-23}~\frac{J}{K} ~\cdot ~300~K	
\end{align*}
The mass of a polystyrene bead can be determined with the density of polystyrene $\rho = 1.05~\frac{t}{m^3}$:
\begin{align*}
m&=\frac{4}{3} \pi \left(\frac{\sigma}{2}\right)^3\cdot \rho\\
&=\frac{4}{3} \pi \left(10^{-6}~m\right)^3\cdot\frac{1}{8}\cdot 1.05~\frac{t}{m^3}\\
&= 0.55 \cdot 10^{-18}~t = 5.5 \cdot 10^{-16}~kg~,\\
\Rightarrow\quad t _{MD}&= 10^{-6}~m
\sqrt{\frac{5.5\cdot 10^{-16}~kgs^2}{3.9\cdot 10^{-21}kgm^2}}\\
&=3.755\cdot 10^{-4}s~.
\end{align*}
$t_{MD}$ should correspond to the structural relaxation time of a bead, i.e., a bead takes roughly $t _{MD}$ to travel to a position over the distance $\sigma$.

An alternative approach to determine relaxation times takes the diffusion into account, which leads to:
\begin{equation}
t = \frac{\sigma ^2}{(6)D}\approx 10^{-2} - 10^{-4}~s
\end{equation}
in corresponding experiments \cite{schoepe}. Therefore, for colloidal particles simulation time scales roughly correspond to experimental time scales.

To simulate Argon, the well depth $\varepsilon = 1.65\cdot 10^{-21}~J$. The diameter of the Argon atoms is equal to $\sigma = 3.4\cdot10^{-10}~m$, the density is $\rho =1.784~ kg m^{-3}$. Inserting in the formula gives:
\begin{equation}
	t=\sqrt{\frac{m\sigma^2}{\varepsilon}}=2.17\cdot10^{-12}~s~.
\end{equation}




\bibliographystyle{plain}
\bibliography{cit} 







\end{small}

\end{document}